\documentclass{amsart}
\numberwithin{equation}{section}

\usepackage{amssymb}
\usepackage[mathscr]{eucal}

\usepackage{times}

\newtheorem{thm}{Theorem}[section]
\newtheorem{lemma}[thm]{Lemma}
\newtheorem{cor}[thm]{Corollary}
\newtheorem{prop}[thm]{Proposition}
\theoremstyle{definition}

\theoremstyle{remark}
\newtheorem{remark}[thm]{Remark}

\renewcommand{\Re}{\mathbb R}

\newcommand{\half}{\frac{1}{2}}
\newcommand{\eps}{\epsilon}

\newcommand{\tr}{\text{\rm tr}}         

\newcommand{\so}{\mathfrak{so}}         

\newcommand{\ii}{\mathbf{i}}

\newcommand{\hGamma}{\hat \Gamma}

\newcommand{\hphi}{\widehat \phi}
\newcommand{\loc}{\text{\rm loc}}
\newcommand{\LL}{\mathcal L}
\newcommand{\MM}{\mathcal M}
\newcommand{\Bo}{\mathcal B}
\newcommand{\Sp}{\mathcal S}
\newcommand{\FF}{\mathcal F}
\newcommand{\oS}{{\mathring{S}}}

\newcommand{\bU}{\bar U}

\newcommand{\Proj}{{\mathbb P}}
\newcommand{\BProj}{\Proj_{\Bo}}
\newcommand{\fProj}{\Proj_{f^{-1}(\Bo)}}
\newcommand{\XX}{\mathbb X}
\newcommand{\YY}{\mathbb Y}
\newcommand{\QQ}{\mathbb Q}

\newcommand{\Id}{\mathbb I}
\newcommand{\fId}{{\Id}_{f^{-1}(\Bo)}}

\newcommand{\kk}{{\mathrm k}}

\title[Static self-gravitating bodies]{Static self-gravitating elastic bodies in Einstein gravity}

\author[L. Andersson]{Lars Andersson} \email{larsa@math.miami.edu}
\address{Albert Einstein Institute, Am M\"uhlenberg 1, D-14476 Potsdam,
  Germany \and
Department of Mathematics, University of Miami, Coral Gables, FL
33124, USA}

\author[R. Beig]{Robert Beig} \email{robert.beig@univie.ac.at}
\address{Physics Faculty/Gravitational Physics, University of Vienna,
Boltzmanngasse 5, A-1090 Vienna, Austria}

\author[B. Schmidt]{Bernd G. Schmidt} \email{bernd@aei.mpg.de}
\address{Max-Planck-Institut f\"ur Gravitationsphysik,
Albert-Einstein-Institut, Am M\"uhlenberg 1, D-14476 Golm, Germany}

\begin{document}
\date{November 20, 2006}



\begin{abstract}
We prove that given a stress-free
elastic body there exists, for sufficiently small values of the
gravitational constant, a unique static
solution of the Einstein equations coupled to
the equations of relativistic elasticity. The solution constructed
is a small deformation of the
relaxed configuration. This result yields the first proof of existence of
static solutions of the Einstein equations without symmetries.
\end{abstract}

\maketitle   






\section{Introduction}
General relativistic effects generated by compact, isolated bodies, such as
stars and even satellites, are of increasing importance in observational
astronomy and experimental general relativity. Considering this fact, it is
remarkable how little is
known about solutions of
the Einstein field equations for systems with spatially compact sources.
The situation is not much better if we describe gravity by Newton's
theory.

The present paper provides the first existence result for
compact, isolated, static elastic bodies in Einstein's theory of
gravity. With the notable exception of collisionless matter,
essentially all previous
results concerning compact, isolated, self-gravitating
bodies deal with static or
stationary fluid bodies. Under reasonable conditions, static fluid bodies are
spherically symmetric, while stationary fluid bodies are axi-symmetric.
Although the variational formulation of elasticity
has strong similarities with that of fluid models, static
elastic bodies may, in contrast to static fluid bodies, be
non-symmetric. In
fact, in this paper we prove, for the first time, existence
of static solutions of the
Einstein equations without symmetries.

\subsection{Compact bodies}
Fluids and dust (i.e. a pressure-less fluid)
are the conceptually simplest and most commonly used matter models. For a
self-gravitating compact body, it is necessary to consider a free boundary
problem with zero traction on the boundary.
%
It was only recently that an existence proof
was given by
Lindblad for the Cauchy problem for a nonrelativistic
perfect fluid with
free boundary, in the absence of gravity \cite{lindblad:incomp,lindblad:comp}. For self-gravitating
fluids, no results of
this generality are known, and it is only in the static or stationary cases
that results are available.

If we assume that spacetime is static, the standard conjecture is
that any isolated self-gravitating body, consisting of a perfect fluid, is
spherically  symmetric. This is known to be the case in
Newton's theory, for a general equation of state. In Einstein's theory there
is the work by Beig and Simon \cite{beig:simon:spherical}
which solves the problem for a large
class equation of states; a proof in the general case remains to be found.
For the case of stationary spacetimes in Einstein gravity,
i.e. spacetimes with a timelike,
non-hypersurface orthogonal Killing field, it is known under certain
additional assumptions on the thermodynamic properties of the fluid that
there exists an additional, rotational Killing field, so that stationary
spacetimes containing fluid bodies are axisymmetric \cite{lindblom:axisym}.

In spite of the symmetry restrictions discussed above, there are
rich classes of stationary and static solutions describing isolated
bodies, even in Newtonian gravity.
%
The almost completely forgotten
work of Leon Lichtenstein from roughly the period 1910-1933,
provides existence results in the Newtonian case for
rotating fluid  solutions in various configurations, see the book
\cite{liechtenstein}.
Inspired by
Lichtenstein, Uwe Heilig showed in 1995
the existence of stationary
rotating fluid solutions in Einstein's theory \cite{heilig}.
%
%

These results allow one to construct
stationary fluid solutions with slowly
rotating almost spherical balls. Further, one has rings, rings around
balls and families of nested rings. It seems almost impossible to get an
overview on all possibilities.
New solutions can often be constructed as
perturbations of known solutions.
For example, starting with a static, spherically symmetric fluid ball
whose existence can be shown by using ODE
techiques, one may prove
the existence
of a rotating solution with small angular velocity. This is essentially what
was done by Lichtenstein and Heilig.

The Vlasov matter model is a statistical
description of weakly interacting particles. It
is conceptually more difficult to work with than fluids but
has  been used very
successfully in various circumstances.
For a survey of known results, see
\cite{rendall:living}.
The existence of various
dynamical and  time independent solution has been demonstrated. All the known
stationary and static solutions have axial symmetry.

\subsection{Elasticity and relativity}
Elasticity is of course one of the oldest topics
of theoretical physics, with origins that can be traced back to the 17'th
century. The book by Marsden and Hughes \cite{hughes:marsden} gives a modern
treatment of elasticity.
Already in 1911, Herglotz \cite{herglotz}
gave a formulation of elasticity in special
relativity.  There are various formulations of elasticity in the
framework of general relativity,
see for example
Rayner \cite{rayner}, Carter and Quintana
\cite{carter:quintana},
Kijowski and
Magli \cite{kijowski:magli:1992,KM},
Christodoulou
\cite{christodoulou:aihp,christodoulou:action},
to name just a few important works.
Strangely enough the problem of existence of static or dynamical
self-gravitating elastic bodies in Einstein's or Newton's theory of
gravity has, to the best of our knowledge, until recently not been
considered. The only exception is for the spherically symmetric case.
Even in non-relativistic elasticity quite little is known. The first
existence theorem in three dimensional
static elasticity was given by Stoppelli in
1954 \cite{stoppelli}.

About six years ago, two of the present authors (R.B. and B.S.), motivated by
this state of affairs, initiated a program to develop existence results for
elasticity in the setting of Einstein's theory of gravity.
%
We first showed the existence of static solutions describing elastic bodies
deformed by their own Newtonian gravitational field \cite{BS:PRS2003}, and
later established the existence of relativistic elastic bodies deformed under
rigid rotation \cite{BS:CQG2005}. For time dependent solutions local in time
and space (no boundary conditions) uniqueness was proved in
\cite{christodoulou:action}, and for existence see
\cite{BS:CQG2003}.  In \cite{BWP} an existence theorem for the motion of  a
free elastic body in special relativity is given.

Elasticity can be described as a Lagrangian field theory
\cite{BS:CQG2003,christodoulou:action}, and hence the action for
self-gravitating elastic bodies is derived by simply adding the gravitational
Lagrangian to the Lagrangian for elasticity.  The  basic matter field is a
map, the configuration, from a region in spacetime to the {\em body}, an
abstract 3--manifold whose points label  the constituents of  the elastic
body moving in spacetime. The stress of a configuration is  determined  by
the {\em stored energy function} which completely fixes the matter model.
This formulation can be used in a non relativistic spacetime, in special
relativity and in Einstein's theory. Elastic materials where the stress is
determined by a stored energy function are usually called {\em hyperelastic}.
Diffeomorphism invariance, a necessary condition in Einstein's theory,
implies that the stored energy function satisfies the additional condition
known in the non-relativistic case as {\em material frame indifference}.

Once the stored energy function is given, the variational problem as well as
the Euler-Lagrange equations are determined.  In particular, one is led to
consider Einstein's field equations with an energy momentum tensor which is
determined by  the deformation. The elasticity equations are a consequence of
the conservation law in
Einstein's theory.

As the configuration is a map from spacetime to the body, we have a free
boundary value problem. To deal with this difficulty, one reformulates
elasticity  using {\em deformations}, i.e. maps from body to spacetime, as
the basic variable. In this setting, known as the material picture, one has a
fixed apriori known boundary.

Let us now consider the static self-gravitating bodies in general relativity.
In this case, the theory can be given a a variational formulation on the
quotient space of the timelike Killing field.
%
%
Thus, in order to construct a static self-gravitating body in Einstein
gravity  we start from a relaxed elastic body without gravitational field and
determine the deformation of such a body under its own gravitational field
for small values of the gravitational constant $G$. To do this, it is
convenient to choose a stored energy function for which there exists a
configuration which is stress free, i.e. which satisfies, together with the
Minkowski metric, the Einstein field equations for $G=0$, as well as the the
elasticity equation. We start from this background solution and construct
nearby self-gravitating solutions for small values of $G$.

As mentioned above, self-gravitating static fluid bodies in general
relativity are known to be spherically symmetric for a large class of
equations of state.  In fact, the result proved in this paper provides the
first example of a static solution to Einstein's field equations which is not
spherically symmetric.

It is worth pointing out that the approach used in this paper cannot be
applied to the problem of constructing for example an elastic neutron star
since in this case, there is no nearby stress free configuration.
To deal with this problem one would have to choose a  stored energy function
in which the shear strains are much smaller  then the hydrostatic
compression. Then one could  begin with a spherically symmetric solution in
which the radial pressure would be different from the tangential
pressure. One could  then use the methods developed  in this paper to
construct nearby  solutions which are not spherically symmetric.

Finally we remark that it may be argued that the result proved in this paper
is very weak, since $G$ is required to be sufficiently small. However, it
should be noted that we make no restriction on the shape  of the body. For
example, one may consider two very large bodies connected  by a very thin
neck. In this situation it is clear that if we make  gravity too strong
(i.e. $G$ too large), the neck will break and hence there can be no static
solution for such a configuration for arbitrarily large values of $G$.  Thus,
without restrictions on the shape  of the undeformed body we can not expect a
stronger result.

\subsection{Overview of this paper}
In section \ref{sec:prel}
we give some analytical preliminaries. We also for the convenience of the
reader review some basic ideas from linearized elasticity which we will make
use of. Further, we prove some results which will be used concerning Bianchi
identities for weakly differentiable metrics, and concerning the divergence
of tensors wich compact support.
Section  \ref{sec:fieldequations}
presents the gravitational field equations in space and the elastic equations
on the body. We use harmonic gauge to  make the field equations elliptic. An
important step is to extend the body to $\Re^3$, the extended body, and to
extend the deformation, the inverse configuration, to a map from the extended
body to physical space, which in our case also is $\Re^3$.  Then we move the
field equations from space to the extended body. In this way we obtain a
quasilinear system of partial differential equations, the reduced Einstein
equations in material form, were the geometrical unknowns are defined on the
extended body and the elastic variables on the body.

In section \ref{sec:analytical}
we formulate the reduced Einstein equations in terms of a non-linear mapping
between Sobolev spaces and calculate the Frechet derivative of the map at the
relaxed configuration. It contains esssentially linearized gravity and
linearized elasticity. The linearized operator is Fredholm with non trivial
kernel and range. The geometric reason for this is the combination of
diffeomorphism invariance of the Einstein equations and Euclidean invariance
of the background solution.
Using an approach to some extent inspired by
\cite{LeDret}, we define a projection operator such that we can use the
implicit function theorem. This way we obtain for small $G$ a solution of the
reduced field equation together with the projected elasticity equations.

Section \ref{sec:bianchi}, which is the heart of the paper,
contains a proof that the solution to the reduced, projected system obtained
using the implicit function theorem, is in fact a solution of the full system
of equations for the self-gravitating elastic body. At first it might seem
that there are two possibilities to prove this.  On the one hand, if the
exact elastic equations were satisfied, a standard argument using the Bianchi
identities would imply that also the harmonicity condition is satisfied and
we have in fact solved the full field equations. On the other hand, if we
could show that the full Einstein equations hold, the Bianchi identities
would imply we also have a solution to the exact elastic equations.

In fact, none of  these two alternatives are applicable, and one must prove
both properties simultaneously. To do this a type of bootstrap argument must
be used.
It is worth mentioning that  the boundary condition of vanishing normal
traction is essential. We could in principle solve the projected elasticity
equation together with the reduced field equations for  a boundary condition
which prescribes the position of  the boundary in space.  However, it would
then in general not be possible to show that all the Einstein field equations
are satisfied. This is consistent with the fact that fixing the surface of a
body in space is not a physical problem in Einstein's theory.

The result of
this paper is proved in a way which is completely different from the
analogous result in our Newtonian paper \cite{BS:PRS2003}. In an appendix
(Appendix \ref{sec:newtonian}) we add an outline of the proof of the Newtonian result, found
and kindly communicated to us by an anonymous referee, which exactly follows
the pattern of the present work.
\section{Preliminaries} \label{sec:prel}
The following index conventions will be used. Upper case latin indices
$A,B,C,\dots$ take values $1,2,3$, lower case latin indices
$i,j,k,\dots$ take values $1,2,3$, and greek indices
$\alpha,\beta,\gamma,\dots$ take values $0,1,2,3$.

We will make use of Sobolev spaces $W^{k,p}$ on domains and the trace spaces
$B^{s,p}$, as well as
weighted Sobolev spaces $W^{k,p}_{\delta}$. Unless there is room for
confusion, the same notation will be used for spaces of tensors and vectors,
as for spaces of scalar functions.
The rest of this section collects
some notions and facts from analysis which shall be needed.
\subsection{Sobolev spaces on domains} \label{sec:sobolev:domain}
The books  \cite{adams:book,burenkov:book} are general references for the
material discussed in this section.
For an integer $k \geq 0$, $1 < p
< \infty$ and a domain $\Omega$,
$W^{k,p}(\Omega)$ is the closure of $C^\infty(\Omega)$ in the norm
$$
||u||_{W^{k,p}(\Omega)} =
\sum_{|\alpha|\leq k} ||\partial^\alpha f||_{L^p(\Omega)}
$$
\def\cprime{$'$}
Further, we shall need the Nikol{\cprime}skii-Besov spaces $B^{s,p} =
B^{s,p}_p$,  which are the
trace spaces for the Sobolev spaces. These are Banach spaces with norm
defined on $\Re^n$ for $s > 0$, $1 < p < \infty$, by
$$
||u||_{B^{s,p}} = ||u||_{L^p} + \left ( \int_{\Re^n} |h|^{-(n+sp)}
||\Delta^{\sigma}_h u||_{L^p}^p dh \right )^{1/p}
$$
where $\sigma$ is the smallest integer strictly greater than $s$,
and $\Delta_h$ is the difference
operator. There are versions of many of the facts stated in this
section also for $p=1$, and $p=\infty$, see the references given
above. Note $B^{k,p} \ne W^{k,p}$, except for the case $p=2$.
Let $k \geq 1$.
A well known fact is that for a domain $\Omega \subset \Re^n$ with $C^k$
boundary, then for
$1 < p < \infty$, the trace $\tr_{\partial \Omega}$ has the property
$$
\tr_{\partial \Omega} W^{k,p} (\Omega) = B^{k-1/p,p}(\partial \Omega)
$$
We shall make use of this fact in the case $k=1$. Further, under these
conditions, there is a bounded linear  extension operator $E:
W^{k,p}(\Omega) \to W^{k,p}(\Re^n) \cap C^\infty (\Re^n \setminus
\bar \Omega)$.
Here $\bar \Omega =
\Omega \cup
\partial \Omega$ denotes the closure of $\Omega$. In fact, for the
last mentioned result to hold, it is sufficient to assume that
$\Omega$ has a Lipschitz regular boundary.

\subsection{The boundary problem of linearized elasticity}
\label{sec:linelast}
The book \cite{valent} is a general reference for the material discussed in
this section.
Let $A_i{}^j{}_k{}^l$
be a fourth order elasticity tensor on a domain $\Omega \subset \Re^n$,
i.e. $A$ has symmetries
$$
A_i{}^j{}_k{}^l = A_j{}^i{}_k{}^l = A_i{}^j{}_l{}^k = A_k{}^l{}_i{}^j
$$
Let
$$
\sigma(u)_i{}^j = A_i{}^j{}_k{}^l \partial_l u^k.
$$
and defined the operator $L$ by
$$
L u_i = \partial_j \sigma(u)_i{}^j
$$
$L$ is strongly elliptic if
$$
A_i{}^j{}_k{}^l \eta^i \xi_j \eta^k \xi_l > 0, \quad \text{ for all
   $\xi,\eta \in \Re^n$} .
$$
For applications in elasticity
it is natural to assume there is a
positive constant $\lambda$ such that for all symmetric
$n\times n$ matrices $\psi^i{}_j$
\begin{equation}\label{eq:Adef}
\lambda |\psi|^2 \leq |A_i{}^j{}_k{}^l \psi^i{}_j \psi^k{}_l| \leq
\lambda^{-1}  |\psi|^2 ,
\end{equation}
see \cite[Chapter III]{valent} or \cite[S 4.3]{hughes:marsden}. The pointwise
stability condition (\ref{eq:Adef}) implies strong ellipticity.
We will assume that (\ref{eq:Adef})
holds for the rest of this section.

The Neumann type problem
\begin{equation}\label{eq:neumann}
L u_i = b_i, \quad \tr_{\partial \Omega} \sigma(u)_i{}^j n_j
= \tau_i
\end{equation}
is equivalent to the statement that
\begin{equation}\label{eq:Aweak}
A(\phi,u) = - \int_\Omega \phi^i b_i + \int_{\partial \Omega} \phi^i \tau_i
\end{equation}
for all $\phi \in C^\infty(\Omega)$,
where $A(u,v)$ is the symmetric bilinear form
$$
A(u,v) = \int_{\Omega} \partial_j v^i A_i{}^j{}_k{}^l \partial_l u^k
$$

Let $H^s(\Omega) = B^{s,2}(\Omega)$.
Then $A$ defines a bounded quadratic form on $H^1(\Omega)$.
The radical $Z$ of $A$ is the space of
$\xi \in H^1(\Omega)$ such that $A(\xi,u)=0$ for all $u \in
H^1(\Omega)$. It follows from (\ref{eq:Adef}) and the symmetry properties of
$A$,
that $Z$
consists of all Euclidean Killing fields $\xi$, i.e. fields of the form
\begin{equation}\label{eq:Killing}
\xi^i = a^i + b^i{}_j x^j, \quad a^i,
b^i{}_j \text{ constants, } b^i{}_j = - b^j{}_i
\end{equation}
Let $H^1(\Omega)_e$ be the $L^2$ orthogonal complement of the radical, i.e.
the space of $u \in H^1(\Omega)$ such that
\begin{equation}\label{eq:uZ}
\int_\Omega \xi_i u^i = 0, \quad \forall \xi \in Z
\end{equation}
Under the above conditions,
the quadratic form $A(u,v)$ is coercive on $H^1(\Omega)_e$.
This follows from the pointwise
stability condition (\ref{eq:Adef}) and Korn's inequality, see
\cite[p. 92]{valent}.
Thus, by the Lax-Milgram theorem, we have that
for any $(b_i, \tau_i) \in H^{-1}(\Omega)
\times H^{-1/2}(\partial \Omega)$
satisfying
\begin{equation}\label{eq:complement}
\int_\Omega \xi^i b_i - \int_{\partial \Omega} \xi^i \tau_i = 0, \quad
\forall \xi \in Z ,
\end{equation}
there is a unique $u \in H^1(\Omega)_e$, which is a weak solution
to (\ref{eq:neumann}), and  which satisfies the estimate
$$
||u||_{H^1} \leq C ( ||b||_{H^{-1}(\Omega)} + ||\tau||_{H^{-1/2}(\partial
  \Omega)} )
$$
We will later refer to (\ref{eq:complement}) as an equilibration condition.
The physical meaning of the equilibration condition is that the total force
and torque exerted by $(b,\tau)$ is zero.
Now, assuming $\partial\Omega \in C^{k+2}$, $A_i{}^j{}_k{}^l \in C^{k+1}$, $u
\in W^{k+2,p}(\Omega)$, one has
from \cite{ADN:II} an estimate of the form
$$
||u||_{W^{k+2,p}(\Omega)} \leq C ( ||L(u)||_{W^{k,p}(\Omega)} +
    ||\tr_{\partial \Omega} \sigma(u)\cdot n||_{B^{1-1/p,p}(\partial \Omega)}
      + ||u||_{L^p(\Omega)} )
$$
Let $W^{k,p}(\Omega)_e$ be the space of $u \in W^{k,p}(\Omega)$ such that
the condition (\ref{eq:uZ}) holds.
Then we have by the above that for
$b,\tau \in W^{k,p}(\Omega)\times B^{k+1-1/p,p}(\partial \Omega)$,
satisfying (\ref{eq:complement}), there is a unique $u \in
  W^{k+2,p}(\Omega)_e$ which solves (\ref{eq:neumann}). In particular, in
  view of the above stated estimates, we have that the linear mapping
$W^{k+2,p}(\Omega) \to W^{k,p}(\Omega) \times B^{k+1-1/p,p}(\partial \Omega)$
defined by
\begin{equation}\label{eq:uop}
u \mapsto (Lu,\tr_{\partial \Omega} \sigma(u)\cdot n)
\end{equation}
is
Fredholm. It follows from the discussion above that the cokernel of the
operator defined by
(\ref{eq:uop}) is
defined by (\ref{eq:complement}), and that the kernel consists
of Killing fields, of the form given in (\ref{eq:Killing}).

\subsection{Weighted Sobolev spaces on $\Re^n$} \label{sec:weighted}
The material which we shall need can be found in \cite{bartnik:mass}.
Let $n > 2$,  $r = |x|$, and
let $\sigma = (1+r^2)^{1/2}$.
For $k \geq 0$, $k$ integer, $1 \leq p \leq \infty$, $\delta \in \Re$,
define function spaces $W^{k,p}_{\delta}$
as the closure of $C^{\infty}_0 (\Re^n)$ in the norms
$$
||u||_{W^{k,p}_{\delta}}^p =
\sum_{|\alpha| \leq k} ||\sigma^{|\alpha|-\delta-n/p} \partial^\alpha u ||_{L^p}^p
$$
We use the notation
$L^p_\delta$ for $W^{0,p}_{\delta}$.
Decreasing $\delta$ means faster decay.
The spaces $W^{k,p}_{s-n/p}$ are equivalent to the
homogenous Sobolev space $\dot W^{k,p}$ without weight, in particular
$L^p_{-n/p} = L^p$. The dual space of $W^{s,p}_{\delta}$ is
$W^{-s,p'}_{\delta'}$ with $1/p + 1/p' = 1$, $\delta' = -\delta -n$.
Weighted H\"older spaces $C^{k,\alpha}_\delta$ can be defined analogously,
see \cite{bartnik:mass}.


Weighted function spaces can be analyzed in terms of ordinary function
spaces by the following standard construction.
Let $\phi$ be a bump function with support in $|x| \leq 2$ and
which equals $1$ on $|x| \leq 1$.
Define a cutoff function $\psi$ by $\psi(x) = \phi(x) - \phi(2x)$. Then
$\psi$ has support in the annulus $1/2 \leq |x| \leq 2$. For a function $u$,
let $u_0 = \phi u$, and let $u_i(x) = \psi(x) u(2^i x)$ be the dilatations of
$u$.
%
%
An equivalent norm for
$W^{k,p}_{\delta}$ is given by
$$
||| u |||_{W^{k,p}_{\delta}}^p =
\sum_{i=0}^\infty 2^{-ip\delta} ||u_i||^p_{W^{k,p}}
$$
%
Using this formulation, inequalities on compact domains may be systematically
generalized to weighted spaces.
The following are some of the basic
inequalities for the weighted spaces.
\begin{enumerate}
\item {\em Inclusion:} If $p_1\leq p_2$, $\delta_2 < \delta_1$, and
$u \in L^{p_2}_{\delta_2}$, then
\begin{equation}\label{eq:includeweighted}
||u||_{L^{p_1}_{\delta_1}} \leq C ||u||_{L^{p_2}_{\delta_2}}
\end{equation}
\item {\em Sobolev I:} If $n-kp > 0$ and $p \leq q \leq np/(n-kp)$, then if
  $u \in W^{k,p}_{\delta}$,
$$
||u||_{L^q_{\delta}} \leq C ||u||_{W^{k,p}_{\delta}}
$$
\item {\em Sobolev II:} If $n < kp$, then if $u \in W^{k,p}_\delta$,
$$
||u||_{L^{\infty}_\delta} \leq C ||u||_{W^{k,p}_{\delta}} ,
$$
and in fact $|u(x)| = o(r^{\delta})$ as $r \to \infty$.
\item {\em Product estimate:}
If $n < kp$, and $u \in W^{k,p}_{\delta_1}$, $v \in
W^{k,p}_{\delta_2}$, then with $\delta = \delta_1 + \delta_2$,
\begin{equation}\label{eq:prodweighted}
||uv||_{W^{k,p}_\delta} \leq C ||u||_{W^{k,p}_{\delta_1}}
||v||_{W^{k,p}_{\delta_2}}
\end{equation}
%
%
\end{enumerate}

Let
$\Delta$ denote the Laplacian defined with respect to the Euclidean metric
on $\Re^n$. We recall some facts about its mapping properties in the setting
of weighted Sobolev spaces. Let $E = \{ j \ :\ j\text{ integer },  j \ne
3-n,\dots,-1\}$. In particular, for $n=3$, $E$ consists of all integers. The
elements of $E$ are called
exceptional weights. A weight $\delta \in \Re \setminus E$
is called nonexceptional. Given $\delta \in \Re$, define  $\kk^-(\delta)$ to
be the largest exceptional weight $< \delta$.

A basic fact is that $\Delta : W^{k,p}_\delta \to
W^{k-2,p}_{\delta-2}$ for $k \geq 2$, $1 < p < \infty$, is Fredholm if and only
if $\delta$ is nonexceptional. In particular, for $\delta \in (2-n, 0)$,
$\Delta : W^{k,p}_\delta \to W^{k-2,p}_{\delta-2}$ is an isomorphism.
%
Let the operator $L$ be of the form $L = a^{ij} \partial_i \partial_j + b^i
\partial_i + c$. Then with $q > n$,
we will say that $L$ is asymptotic to $\Delta$, of order $\tau < 0$, if
$$
a^{ij} - \delta^{ij} \in W^{2,q}_{\tau}, \quad b^i \in W^{1,q}_{\tau-1}, \quad
c \in L^{q}_{\tau-2}
$$
Note that the above conditions are stronger than those stated in
\cite[Definition 1.5]{bartnik:mass}, and that we use
the opposite sign convention for $\tau$.
If $L$ is asymptotic to $\Delta$, then for $1 < p \leq q$,
$L: W^{2,p}_{\delta} \to L^p_{\delta}$
is bounded.

The following version of elliptic regularity is easily proved using standard
estimates for elliptic operators on domains (see for example \cite[Chapters
  8,9]{GT}) and scaling.
Suppose $L$ is asymptotic to $\Delta$ of order $\tau < 0$ and suppose $2 \leq  p
\leq q$.
For $u \in W^{1,p}_\delta$, such that $Lu \in
L^p_{\delta-2}$, elliptic regularity gives $u \in
W^{2,p}_{\delta}$ and the inequality
$$
||u||_{W^{2,p}_{\delta}} \leq C ( ||Lu||_{L^p_{\delta-2}} + ||u||_{L^p_\delta} )
$$
holds.
See \cite[Proposition 1.6]{bartnik:mass} for a stronger version of elliptic
regularity.

If $\delta$ is nonexceptional, there are constants $C,R$ so that
if $u \in W^{1,p}_{\delta}$,
the scale broken estimate
$$
||u||_{W^{2,p}_{\delta}} \leq C ( ||Lu||_{L^p_{\delta-2}} + ||u||_{L^p(B_R)} )
$$
holds, cf. \cite[Theorem 1.10]{bartnik:mass}.
Here $B_R = \{x\ : \ |x|\leq R\}$. We will now state a consequence of
this estimate which we shall make use of.
Assume that
$\delta+\tau$ is non-exceptional. If $ u \in
W^{2,p}_{\delta}$, $Lu \in L^p_{\delta+\tau - 2}$, then
for exceptional values $\delta+\tau < j \leq \kk^-(\delta)$,
there are $h^j \in C^\infty(\Re^n)$, harmonic and homogenous of order $j$ in
$\Re^n \setminus B_R$, such that
$$
u = \sum_{\delta+\tau < j \leq \kk^-(\delta)} h^j + v
$$
where $v \in W^{2,p}_{\delta+\tau}$, and an estimate of the form
$$
\sum_j ||h^j||_j + ||v||_{W^{2,p}_{\delta+\tau}} \leq C ( ||Lu||_{L^p_{\delta+\tau-2}} +
    ||v||_{B_R} )
$$
holds.
Here $||h^j||_j$ is a suitable norm of for homogenous, harmonic functions, for
example $||h^j||_j = ||r^{-j} h^j||_{L^\infty(S_{2R})}$, with $S_{2R} = \{x\ : \
|x|=2R\}$.
In particular, if $\ker L = 0$ on $W^{2,p}_{\delta}$, then the above estimate
takes the form
$$
\sum_j ||h^j||_j + ||v||_{W^{2,p}_{\delta+\tau}} \leq C
||Lu||_{L^p_{\delta+\tau-2}}
$$
To make the above explicit, suppose $n=3$, $-1 < \delta < 0$ and $-2 < \delta
+ \tau < -1$. Then $\kk^-(\delta)=-1$ and with $u \in W^{2,p}_\delta$, $Lu \in
L^p_{\delta+\tau-2}$, we have
$$
u = \frac{c_1}{r} \zeta + v
$$
where $\zeta$ is a cutoff function such that $\zeta=1$ in $\Re^3 \setminus
B_R$,and $\zeta = 0$ in $B_{R/2}$,  and where
$v \in W^{2,p}_{\delta+\tau}$ satisfies an estimate of the form
$$
|c_1| + ||v||_{W^{2,p}_{\delta+\tau}} \leq C ( ||Lu||_{L^p_{\delta+\tau-2}} +
||v||_{B_R} )
$$
We shall make use of this estimate in section \ref{sec:eulerian}.

For $\tau < 0$, $p > n$,
define the space
$E^{k,p}_{\tau}$ of asymptotically Euclidean metrics on $\Re^n$ as the
space of $h_{ij}$ such that
$$
h_{ij} - \delta_{ij} \in W^{k,p}_{\tau}
$$
where $\delta_{ij}$ denotes the flat Euclidean metric on $\Re^n$. Then
$E^{k,p}_{\tau}$ is a Banach manifold.

Let
$R_{ij}$ be the Ricci tensor of $h_{ij}$.
We shall
make use of the fact that if $h \in E^{2,p}_{\tau}$
for $p > n$, the operators $\Delta_h$ and
$V^i \mapsto LV_i = \Delta_h V_i + R_{ij} V^j$ are asymptotic to $\Delta$ of
order $\tau$. Here it should be noted that the principal part of the
operator $L$ is the scalar Laplacian, acting diagonally.
The results concerning elliptic operators that we have stated in this
subsection generalize immediately to the case of elliptic systems of diagonal
form.

\subsection{Bianchi identity}
The Bianchi identity for weakly regular Riemann spaces will
play an important role in this paper,
and therefore we give a proof of this fact below. The considerations in this
subsection are local, and we work in local
Sobolev spaces, denoted by $W^{k,p}_{\loc}$. By definition, $f \in
W^{k,p}_{\loc}$ if for each compact domain $U \subset M$, $f \in
W^{k,p}(U)$.
For $1 < p < \infty$,
let $q$ be the dual exponent of $p$, such that $1/p + 1/q = 1$.
For non-negative
integers $k$, we define
$W^{-k,q}_{\loc}$ as the dual space to $W^{k,p}_{\loc}$. Then
$\partial^{\alpha} f \in W^{-k,p}_{\loc}$ if $f \in L^p_{\loc}$, where $k =
|\alpha|$.

We make note of the following product estimates. Suppose that $p > n$. If $u
\in W^{2,p}_{\loc}$, $v \in W^{1,q}_{\loc}$, then $uv \in W^{1,q}_{\loc}$. To
see this, differentiate the product and use Sobolev imbedding.
Further, if $u \in
W^{2,p}_{\loc}$, $w \in W^{-1,p}_{\loc}$, then $uw \in W^{-1,p}_{\loc}$. We
estimate the product $uw$ as follows. Let $v \in W^{1,q}_{\loc}$ and consider
for any domain $U$ with compact closure, $ \int_U vuw \,. $  By the above
mentioned estimates, $vu \in W^{1,q}(U)$.  Thus the integral is well defined,
and the inequality  $|\int_U vuw| \leq C || v ||_{W^{1,q}(U)} || u
||_{W^{2,p}(U)} ||w||_{W^{-1,p}(U)}$ holds.    But $v \in W^{1,q}(U)$ was
arbitrary. It follows that $uw$ defines a bounded linear functional on
$W^{1,q}(U)$, and hence  $uw \in W^{-1,p}_{\loc}$.

\begin{lemma}[Bianchi identities] \label{lem:bianchi}
Consider a Riemann manifold $(M, h_{ij})$ of dimension $n$, with metric
 $h_{ij} \in W^{2,p}_{\loc}$, $p > n$.  Then the first Bianchi identity holds
 for $R_{ijkl}$.  Further, the second Bianchi identity
$$ \nabla_{[m} R_{ij]kl} = 0
$$ and the contracted second Bianchi identity
$$ \nabla^i R_{ij} - \half \nabla_j R  = 0
$$ are valid in the sense of distributions.
\end{lemma}
\begin{proof}
The first two statements are clear from the product estimates and the
definition of $\nabla$ and $R_{ijkl}$. It is most convenient to prove the
Bianchi identity using the Cartan formalism. Let $\theta$ be an orthornomal
coframe. The structure equations are
\begin{align*}
d\theta + \omega \wedge \theta &= 0 \\ d\omega + \omega \wedge \omega &=
\Omega
\end{align*}
Assuming $h_{ij} \in W^{2,p}_{\loc}$ we have $\omega \in W^{1,p}_{\loc}$ and
$\Omega \in L^p_{\loc}$.

The Bianchi identity is the statement $d^{\omega} d^{\omega} d^{\omega} = 0$,
where $d^{\omega}$ is the covariant exterior derivative. Recall that on a
section of a tensor bundle, $d^\omega d^\omega s= (d\omega + \half [\omega
\wedge \omega]) s = \Omega s$ and on a $\so(n)$-valued tensor, such as
$\Omega$, $d^\omega H = dH + [\omega\wedge H]$. Evaluating $d^\omega d^\omega
d^\omega$ gives
\begin{align*}
d^\omega \Omega &= d\Omega + [\omega \wedge \Omega] \\ &= d^2 \omega + \half
d[\omega \wedge \omega] + [\omega \wedge \Omega]
\end{align*}
Now, $d^2 = 0$ on distributions. Further, expanding the other terms in the
right hand side gives products of elements of $W^{1,p}_{\loc}$ and
$L^p_{\loc}$. Therefore the standard algebraic identities hold to show that
$d^\omega \Omega = 0$ in the sense of distributions.  This is equivalent to
the statement that $\nabla_{[m} R_{kl]i}^{\ \ \ j} = 0$ in the sense of
distributions.  Contracting this identity twice gives by the standard
argument (making use of the product estimates stated above, and the fact that
$\nabla_m R_{kli}{}^j \in W^{-1,p}_{\loc}$, to justify the contraction), the
identity
$$ \nabla^i R_{ij} - \half \nabla_j R = 0
$$ which holds in the sense of distributions.
\end{proof}

For a domain $\Omega$ with boundary $\partial \Omega$, let $\chi_\Omega$
denote the characteristic function of $\Omega$, and $\tr_{\partial \Omega}$
the trace to the boundary.
The following Lemma characterizes the divergence free tensors supported on a
domain.
\begin{lemma} \label{lem:div}
Let $(M, h_{ij})$ be a Riemann manifold of dimension $n$ with metric of class
  $W^{2,p}_{\loc}$, $p > n$.  Let $\Omega$ be a bounded domain compactly
  contained in $M$.  Assume that $\Omega$ has $C^1$ boundary $\partial
  \Omega$, and let $T_{ij}$ be a symmetric tensor of class $W^{1,p}_{\loc}$.
Then
$$
\nabla^i (T_{ij} \chi_\Omega ) = (\nabla^i T_{ij}) \chi_\Omega \in L^p
$$
%
if and only if the zero traction condition
$$ (\tr_{\partial \Omega} T_{ij}) n^j  = 0
$$
holds, where $n^j$ denotes the normal of $\partial \Omega$.
In particular, the identity
$$ \nabla^i (T_{ij} \chi_\Omega) = 0
$$ holds in the sense of distributions, if and only if $(\nabla^i T_{ij} )
\chi_\Omega = 0$ and the zero traction condition holds.
\end{lemma}
\begin{proof}
Let $Y^i \in C^\infty_0$. Then we have
\begin{align*}
\int_M  Y^j \nabla^i (T_{ij} \chi_\Omega) &= - \int_M \nabla^i Y^j T_{ij}  \\
&= - \int_\Omega \nabla^i Y^j T{ij} \\ &= \int_\Omega Y^j \nabla^i T_{ij} +
\int_{\partial \Omega} Y^j T_{ij} n^i
\end{align*}
where $n^i$ is the outward normal of $\partial \Omega$.  This implies that
with $1/p + 1/q = 1$, the inequality
$$ |\int_M Y^j \nabla^i (T_{ij} \chi_\Omega) | \leq ||Y||_{L^q} ||(\nabla^i
T_{ij}) \chi_\Omega ||_{L^p}
$$ holds if and only if $(\tr_{\partial \Omega} T^{ij}) n^j = 0$. This proves
the Lemma.
\end{proof}


\section{The field equations for a static, self-gravitating elastic body}
\label{sec:fieldequations}
We first consider a variational formulation of a self-gravitating elastic
body in a 3+1 dimensional spacetime $(\MM, g_{\alpha\beta})$ and then
specialize this to the static case.  The {\em body} $\Bo$ is a 3-manifold,
possibly with boundary. We use coordinates $X^A$ on $\Bo$, and $x^{\alpha}$
on spacetime.
In the Eulerian formulation of elasticity, the body is described by {\em
configurations} $f: \MM \to \Bo$.
%
The total Lagrangian density for the Einstein-matter system under
consideration is, setting the speed of light $c=1$ for convenience,
$$ \LL = - \frac{1}{16\pi G} \sqrt{-g} R_g + \sqrt{-g} \rho
$$ where $\rho = \rho(f, \partial f, g)$ is the energy density of the
materical in its own rest frame.  General covariance implies that $\rho$ is
of the form $\rho = \rho(f^A, \gamma^{AB})$ where $\gamma^{AB} =
f^A{}_{,\alpha} f^B{}_{,\beta} g^{\alpha\beta}$, where $f^A_{,\alpha} =
\partial_{\alpha} f^A$.
%
Geometrically, this  means the
following: there is a function $\hat{\rho}$ on
the bundle of contravariant, symmetric two-tensors over
$\mathcal{B}$, and $\rho (f^A, \gamma^{AB})$ is the composition
$\hat{\rho} \circ f_* (g^{-1})$, where $g^{-1}$ is the inverse
metric and $f_*$ is pushforward under the map $f$, acting on
contravariant two-tensors. For an equivalent description of elastic
materials see \cite{christodoulou:aihp,christodoulou:action}. 
These references require in addition that
$\mathcal{B}$ be furnished with a crystalline structure, which is
essentially a choice of three linearly independent vector fields on
$\mathcal{B }$ describing (the continuum limit of) the crystal
lattice. More precisely, a crystalline structure is a linear
subspace of the space of vector fields on $\mathcal{B }$ with the
following property: for all points of $\mathcal{B }$ the evaluation
map, which sends vector fields on $\mathcal{B }$ to tangent vectors
at this point of $\mathcal{B }$, is an isomorphism, when restricted
to this subspace. 
Our assumptions, in section \ref{sec:constitute} below, will render such
a choice of vector fields which Lie commute. This means there are no
no dislocations in the crystal lattice.
%

We now specialize to the static case. Let $\MM = \Re \times M$, where
the {\em space manifold} $M$ is diffeomorphic to Eucidean 3-space,
$M \cong \Re^3_{\Sp}$. Further, we take the body $\Bo$ to be a bounded open
domain in Eucliden 3-space, $\Bo \subset \Re^3_{\Bo}$. We refer to
$\Re^3_{\Bo}$ as the {\em extended body}, and will use coordinate $X^A$ also
on $\Re^3_{\Bo}$. The body $\Bo$ will be assumed to have smooth boundary
$\partial \Bo$, and the closure $\Bo \cup \partial  \Bo$ will be denoted by
$\bar \Bo$. Letting $(x^{\alpha}) = (t, x^i)$ where $x^i$ are coordinates on
$M$, we can write the static spacetime metric in the form
\begin{equation}\label{eq:ds2}
g_{\alpha\beta} dx^\alpha dx^\beta = - e^{2 U} dt^2 + e^{-2  U}
h_{ij} dx^i dx^j
\end{equation}
where $U, h_{ij}$ depend only on $x^i$.
Further, the configurations
$f : M \to \Bo$ are assumed to depend only on $x^i$. Assuming that a volume
form $V_{ABC}$,  on $\Bo$ is given, we may
introduce the particle number density $n$ by
\begin{equation}\label{n}
f^A{}_{,i}(x) f^B{}_{,j}(x) f^C{}_{,k}(x) V_{ABC}(f(x)) = n (x)
\,\epsilon_{ijk}(x)
\end{equation}
where $\epsilon_{ijk}$ is the volume element of $h_{ij}$. Note that
the actual number density defined with respect to the metric
$e^{-2U} h_{ij}$ is $e^{3U} n$.
We assume the configurations $f$ are orientation preserving in the
sense that $n$ is positive on $f^{-1}(\Bo)$.

Let
\begin{equation}\label{eq:HABdef}
H^{AB} = f^A{}_{,i} f^B{}_{,j} h^{ij}
\end{equation}
so that
$\gamma^{AB} = e^{2U} H^{AB}$. Note that equation (\ref{n})
implies
$$
6 \,n^2 = H^{AA'} H^{BB'}H^{CC'}V_{ABC} V_{A'B'C'}
$$
The Lagrangian density
$\LL$ is in terms of these variables, modulo a total divergence,
\begin{equation}\label{eq:Lstat}
\LL = - \frac{1}{16\pi G}\sqrt{h} (R - 2 |\nabla U|^2) + e^{ U} n \eps \sqrt{h}
\end{equation}
Here, $R$ is the scalar curvature of $h_{ij}$, $|\nabla U|^2 = h^{ij} \nabla_i U \nabla_j
U$, and the {\em relativistic stored energy function} $\eps$, defined by $\rho = n \eps$,
is of the form $\eps =
\eps(f^A, e^{2 U} H^{AB})$, where $\eps$ is a smooth function of its
arguments.
In particular, by the chain rule,
we have $\partial \eps/\partial H^{AB} = e^{2 U} \partial \eps/\partial
\gamma^{AB}$.

Suppose that a non-relativistic stored energy function $w(f^A , K^{AB})$ is
given, for example one suggested by experiment, where $K^{AB}$ is the
non-relativistic analogue of $H^{AB}$. A relativistic stored energy
function $\eps$ corresponding to $w$ can be defined
as a sum of the {\em specific
rest mass} $\mathring{\eps}$ and the relativistic analogue
$w(f^A, e^{2U} H^{AB})$ of the stored energy function.
The specific rest
mass is defined such that $\mathring{\eps} V_{ABC}$ is the rest mass distribution of
the material in its natural state. It can be shown that
if the dependence on the light
speed $c$ taken properly into account,
the field equations tend to those of the corresponding Newtonian
model when we let $c \nearrow \infty$. See \cite{BS:CQG2003} for details.

\subsection{Field equations in Eulerian form} \label{sec:field-eulerian}
In order to write the field equations, we introduce the stress tensor
$\sigma$. We will need the form of the stress tensor on the body and in
space. These are given by
\begin{equation}\label{stress}
\sigma_{AB}=-2 \frac{\partial \epsilon}{\partial H^{AB}}, \quad \sigma_{ij}=
n f^A{}_{,i}f^B{}_{,j}\, \sigma_{AB}, \quad \sigma_i{}^A=f^B{}_{,i}\,
\sigma_{BC} H^{CA}
\end{equation}
We remark that our convention for the stress here is that used in standard
nonrelativistic elasticity, as opposed to the usual one in general
relativity.  
It is important to note
that the elastic quantities such as $H^{AB}$, $\sigma_{ij}$, viewed as
functions on space, are only defined on $f^{-1}(\mathcal{B})$. The
Euler-Lagrange equations resulting from the Lagrangian (\ref{eq:Lstat}) are
\begin{subequations} \label{eq:Euler}
\begin{align}
\nabla_j (e^U \sigma_i{}^j) &= e^U (n \epsilon - \sigma_l{}^l) \nabla_i U
\quad \text{\rm in } f^{-1}(\Bo), \quad \sigma_i{}^j n_j |_{f^{-1}(\partial
\Bo)}=0 \label{elast} \\ \Delta_h U &= 4 \pi G e^U(n \epsilon - \sigma_l{}^l)
\chi_{f^{-1}(\mathcal{B})}\quad \text{\rm in } \Re^3_{\Sp}
\label{potential} \\
G_{ij} &= 8 \pi G ( \Theta_{ij} - e^U
\sigma_{ij}\,\chi_{f^{-1}(\mathcal{B})})\quad \text{\rm in } \Re^3_{\Sp}
\label{metric}
\end{align}
\end{subequations}
where
\begin{equation} \label{theta}
\Theta_{ij} = \frac{1}{4 \pi G}[\nabla_i U \nabla_j U - \frac{1}{2} h_{ij}
|\nabla U|^2].
\end{equation}
The system (\ref{eq:Euler}) is equivalent to the 4-dimensional Einstein
equations
\begin{equation}\label{4}
G_{\mu \nu}= 8 \pi G T_{\mu \nu},
\end{equation}
where $G_{\mu \nu}$ is the Einstein tensor of the static Lorentz metric given
by (\ref{eq:ds2}) on $\Re \times M$, and
\begin{equation}\label{Tmunu}
T_{\mu\nu} dx^\mu dx^\nu = e^U [e^{2U} n\, \epsilon \, (u_\mu dx^\mu)^2 -
\sigma_{ij}\, dx^i dx^j]\,,\;\;\;u^\mu \partial_\mu = e^{-U}\partial_t
\end{equation}

The equations (\ref{potential},\ref{metric}) together imply the elasticity
equation (\ref{elast}). The reason is that the contracted Bianchi identity
for $G_{ij}$, $\nabla^i G_{ij} = 0$, implies that the right hand side of
(\ref{metric}) has vanishing divergence, and in particular the divergence of
the compactly supported term $e^U \sigma_{ij} \chi_{f^{-1}(\Bo)}$ must be
well defined.  By Lemma \ref{lem:div} this implies that the zero traction
boundary condition in equation (\ref{elast}) holds, and hence by equations
(\ref{potential}) and (\ref{theta}), equation (\ref{elast}) follows.

Let $\hat{\delta}$ be a fixed background metric on $M$, which we will take to
be Euclidean, and let $\hGamma^i_{jk}$ be the Christoffel symbol of
$\hat{\delta}$. Then, with
\begin{equation}\label{eq:Vdef}
V^i = h^{jk}(\Gamma^i_{jk} - \hat{\Gamma}^i_{jk}),
\end{equation}
$-V$ is the tension field of the identity map $(M, h_{ij}) \to (M,
\hat{\delta})$, and we have the identity
\begin{equation}\label{AM}
R_{ij}=-\frac{1}{2}\Delta_h h_{ij} + \nabla_{(i}V_{j)} + Q_{ij}(h,\partial h),
\end{equation}
where $\Delta_h h_{ij}$ is the scalar Laplacian defined with respect to
$h_{ij}$, acting on the components of $h_{ij}$ and $Q_{ij}$ is quadratic in
$\partial h$. We use the notation $t_{(ij)} = \half (t_{ij} + t_{ji})$ for
the symmetrization of a tensor.  In particular, $h_{ij} \mapsto R_{ij} -
\nabla_{(i}V_{j)}$ is a quasilinear elliptic operator, while $h_{ij} \mapsto
R_{ij}$ fails to be elliptic. This failure is essentially due to the
covariance of $R_{ij}$. It follows that also the system (\ref{eq:Euler})
fails to be elliptic. In order to construct solutions to (\ref{eq:Euler}), we
will replace equation (\ref{metric}) by the reduced system which results from
replacing $R_{ij}$ by $R_{ij} -  \nabla_{(i}V_{j)}$. The modified system
which we will consider is of the form
\begin{equation}\label{reduced}
-\frac{1}{2}\Delta_h h_{ij} + Q_{ij}(h,\partial h) = 2 \nabla_i U \nabla_j U
- 8 \pi G e^U (\sigma_{ij} - h_{ij}\,\sigma_l{}^l)\chi_{f^{-1}(\mathcal{B})}
\end{equation}

\subsection{Field equations in material form}
\label{sec:material}
In the Eulerian formulation above, the elasticity equation (\ref{elast}) is a
nonlinear system with Neumann type boundary conditions on the domain
$f^{-1}(\Bo)$ which depends on the unknown configuration $f$. We will avoid
dealing directly with this ``free boundary'' aspect of the system
(\ref{eq:Euler}) by passing to the material, or Lagrangian form of the
system.  In this picture the configurations $f : M \to \Bo$ are replaced by
{\em deformations}, i.e maps $\phi: \Bo \to M$ satisfying $\phi = f^{-1}$ on
$\Bo$. Recall that the body $\Bo$ is a bounded open domain in $\Re_{\Bo}$,
the extended body, and that $\Bo$ is assumed to have a smooth boundary
$\partial \Bo$. We will assume throughout the rest of this paper that $\Bo$
is connected. See remark \ref{rem:connected} for discussion on this point.

We assume given a diffeomorphism $\ii : \Re^3_{\Bo} \to \Re^3_{\Sp}$, which
is an isometry, $\ii^* \hat\delta = \delta_{\Bo}$.  In Cartesian coordinate
systems $X^A$ on $\Re^3_{\Bo}$ and $x^i$ on $\Re^3_{\Sp}$ where
$\delta_{\Bo}$ and $\hat \delta$ have components $\delta_{AB}$ and
$\delta_{ij}$ respectively, $\ii$ can be assumed to have the form $\ii^i(X) =
\delta^i_A X^A$, so that $\partial_A \ii^i = \delta^i_A$.
Since $\Bo$ has smooth boundary, functions on $\Bo$ can be extended to the
whole space, and in particular, given $\phi$, there is an extension $\hphi:
\Re^3_{\Bo} \to \Re^3_{\Sp}$, depending smoothly on $\phi$, such that
$\hphi(X) = \ii(X)$ for $X$ outside some large ball.

In the material picture, the dependent variables $f, U, h_{ij}$ are replaced
by the fields $\phi, \bU, \overline{h_{ij}}$ which we now introduce. As
mentioned above, $\phi$ is assumed to be a  diffeomorphism $\Bo \to \phi(\Bo)
\subset \Re^3_{\Sp}$, and the extension $\hphi$ of $\phi$, which depends real
analytically on $\phi$, is used to define the fields $\bU = U \circ \hphi$, a
function on $\Re^3_{\Bo}$, and $\overline{h_{ij}} = h_{ij} \circ \hphi$, a
metric on $\Re^3_{\Bo}$.
We will use the same symbols for these fields restricted to $\Bo$.
\begin{remark}
It is important to note that $\overline{h_{ij}} \ne \hphi^* h_{ij}$, since we
are only pulling back the components of $h_{ij}$ in the coordinate system
$(x^i)$, not the tensor itself. In particular, $\overline{h_{ij}}$ does not
transform as a tensor and is more propertly viewed as a collection of
scalars. See appendix \ref{sec:bobbyapp} for discussion.
\end{remark}
The equation (\ref{n}) defining $n$ can be written in the form $f^* V = n
\mu_h$. where $\mu_h$ is the volume element of $h$. Defining $J = n^{-1}$, we
have $\phi^* \mu_h = JV$. The Piola transform of $\sigma_i{}^j$ can now be
written in the form
$$ \bar \sigma_i{}^A = J (f^A{}_{,j} \sigma_i{}^j) \circ \phi
$$ With this notation we have in particular the relation $\nabla_A \bar
\sigma_i{}^A = J (\nabla_j \sigma_i{}^j)\circ \phi$.  To derive the material
version of (\ref{elast}) one may use this relation directly, or proceed by
first pulling back the matter Lagrangian to $\Bo$ and then applying the
variational principle, see \cite{BS:CQG2005}. One finds
\begin{equation}\label{elastmaterial}
\nabla_A(e^{\bar{U}} \bar{\sigma}_i{}^A)= e^{\bU} [\bar{\epsilon} -
   \frac{\bar{\sigma}_l{}^l}{\bar{n}}]\,\overline{\partial_i U}\quad
   \text{\rm in } \mathcal{B},\quad \bar{\sigma}_i{}^A
   n_A|_{\partial\mathcal{B}}=0
\end{equation}
Here $\nabla_A (e^{\bU} \bar \sigma_i{}^A)$ is defined in terms of the volume
element $V$ and does not involve a choice of metric on $\Bo$. We have
$$ \nabla_A(e^{\bU} \bar{\sigma}_i{}^A)= \frac{1}{V} \partial_A(V e^{\bU}
\bar{\sigma}_i{}^A)-e^{\bU} \phi^j{}_{,A} \overline{\Gamma^k_{ij}}
\bar{\sigma}_k{}^A
$$ The bars in Eq.(\ref{elastmaterial}) correspond to the convention that
$f^A{}_{,i}$ be replaced by $\psi^A{}_i$ defined as a functional of $\phi$ by
\begin{equation}\label{psi}
\psi^A{}_i(X) \hphi^i{}_{,B}(X)=\delta^A{}_B
\end{equation}
and $H^{AB}$ be changed into
\begin{equation}\label{Hnew}
\bar{H}^{AB}= \psi^A{}_i \psi^B{}_j \overline{h^{ij}},
\end{equation}
thus $\bar{H}^{AB}$ is the inverse of $\hphi^* h_{ij}$. Note $\psi^A{}_i$ is
defined on $\Re^3_{\Bo}$.  With $\bar{\epsilon} = \bar{\epsilon}(X,e^{2
\bar{U}}\bar{H}^{AB})$ and $\bar{H}^{AB}$ understood in this sense, we have
the identity
\begin{equation}\label{Piola}
\bar{\sigma}_i{}^A = \frac{\partial \bar{\epsilon}}{\partial \phi^i{}_{,A}}.
\end{equation}
In particular, $\bar \sigma_i{}^A = \psi^B{}_i \bar \sigma_{BC} \bar H^{CA}$,
and hence (\ref{elastmaterial}) is a second order equation for $\phi$.  For
barred quantities the corresponding rule for bars of derivatives gives for $U$
\begin{equation}\label{over}
\overline{\partial_i U}= \psi^A{}_i\, \partial_A \overline{U}
\end{equation}
For equations (\ref{potential}) and (\ref{reduced}) we simply replace each
term by its barred version, i.e.
\begin{equation}\label{barU}
\overline{\Delta_h U} = 4 \pi G e^{\bar{U}}(\bar{n} \bar{\epsilon} -
\bar{\sigma}_l{}^l) \chi_{\mathcal{B}}\quad \text{\rm in } \Re^3_{\Bo}
\end{equation}
Further, the covariance of the Laplacian implies $\overline{\Delta_h U} =
(\Delta_h U)\circ \hphi = \Delta_{\hphi^* h} (U \circ \hphi)$, so that in
this expression, the pullback $\hphi^* h$ appears. This tensor is given by
the inverse of $\bar H^{AB}$. It follows that $\overline{\Delta_h U}$
involves second derivatives of $\hphi$.  An analogous remark applies to
\begin{equation}\label{barreduced}
-\frac{1}{2}\overline{\Delta_h h_{ij}} +
Q_{ij}(\overline{h},\overline{\partial h}) = 2 \overline{(\nabla_iU)}
\overline{(\nabla_j U)} - 8 \pi G e^{\bar{U}} (\bar{\sigma}_{ij} -
\overline{h_{ij}}\,\bar{\sigma}_l{}^l)\chi_{\mathcal{B}}.
\end{equation}

\subsection{Constitutive conditions} \label{sec:constitute}
We shall need to assume that the elastic
material is able to relax in Euclidean space (which in particular implies the
absence of dislocations, see the remark at the beginning of section
\ref{sec:fieldequations}).  Further, the stored energy function $\epsilon$
must be such that the linearized elasticity operator is elliptic. In fact, we
shall assume the pointwise stability condition. The constitutive conditions
for $\epsilon$ are thus formulated as follows. There should exist a Euclidean
metric $\delta_{\mathcal{B}}=\delta_{AB}\,dX^A dX^B$ on $\mathcal{B}$ (we
will use the same symbol for this metric extended to
$\mathbb{R}^3_{\mathcal{B}}$) such that
\begin{equation}\label{const}
\mathring{\epsilon}(X)=\epsilon |_{(U=0,H = \delta_{\mathcal{B}})}\geq C,
\quad \left( \frac{\partial \epsilon}{\partial H^{AB}}\right)|_{(U=0,H =
\delta_{\mathcal{B}})}=0\;\;\;\mathrm{in}\;\mathcal{B}
\end{equation}
and
\begin{equation} \label{pointwise}
\mathring{L}_{ABCD}\, N^{AB}N^{CD} \geq C'(\delta_{CA}\delta_{BD}+
\delta_{CB}\delta_{AD})\,N^{AB}N^{CD}\;\;\;\mathrm{in}\;\mathcal{B},
\end{equation}
where
\begin{equation}\label{second}
\mathring{L}_{ABCD}(X) := \left(\frac{\partial^2 \epsilon}{\partial
H^{AB}\partial H^{CD}}\right)|_{(U=0,H=\delta_{\mathcal{B}})}\;\;.
\end{equation}
and $C,\,C'$ are positive constants.  The condition (\ref{pointwise}) is just
the pointwise stability condition (\ref{eq:Adef}) discussed in section
\ref{sec:linelast}.
The quantity $\mathring{\epsilon}$ appearing in (\ref{const}) is the rest
mass term in the relativistic stored energy function, as discussed above.
For physical reasons, and in fact for hyperbolicity of the time dependent
theory, it is necessary to assume that $C$ is positive.  However, for the
purposes of the present paper, this condition could be dropped.  We shall
assume, for simplicity, that $V_{ABC}$ is the volume form associated with
$\delta_{\mathcal{B}}$ (i.e.  that $V_{123}=1$ in Euclidean coordinates), so
that $n \sqrt{h} = det(\partial f)$.

\section{Analytical setting} \label{sec:analytical}
We will use the implicit function theorem to construct, for small values of
Newton's constant $G$,  static self-gravitating elastic bodies near the
reference state described in section \ref{sec:constitute}. We will use the
field equations in the material form given by (\ref{elastmaterial}),
(\ref{barU}) and (\ref{barreduced}).
%

We will now introduce the analytical setting where this work will be carried
out.
Fix a weight $\delta \in (-1,1/2)$.  This choice of $\delta$ determines the
weighted Sobolev spaces which will be used in the implicit function theorem
argument below. The range of weights for which the isomorphism property for
$\Delta$ holds is $(-1,0)$ but we shall need $\delta \in (-1,1/2)$ later
on. Further, we fix $p > 3$ to be used in setting up the function spaces
which will appear in our argument.
The body $\Bo$ is a bounded open domain in $\Re_{\Bo}$, the extended body,
with smooth boundary $\partial \Bo$. Under these conditions, the trace to the
boundary $\tr_{\partial\Bo}$ is a continuous linear map $W^{1,p}(\Bo) \to
B^{1-1/p,p}(\partial \Bo)$, and there is a bounded, linear extension operator
$E: W^{2,p}(\Bo) \to W^{2,p}_\loc(\Re^3_{\Bo})$, see the discussion in
section \ref{sec:sobolev:domain} or \cite{adams:book}.
The spaces which will be used in the implicit function theorem argument are
$B_1 = W^{2,p}(\Bo) \times W^{2,p}_{\delta} \times E^{2,p}_\delta$, and let
$B_2 = [L^p(\Bo) \times B^{1-1/p,p}(\partial \Bo)] \times L^p_{\delta-2}
\times L^p_{\delta-2}$. Thus, $B_1$ is a Banach manifold, and $B_2$ is a
Banach space.

The residuals of equations (\ref{elastmaterial}), (\ref{barU}) and
(\ref{barreduced}) define a map $\FF : \Re \times B_1 \to B_2$, $\FF = \FF(G,
Z)$, where we use $Z = (\phi, \bU, \overline{h_{ij}})$ to denote a general
element of $B_1$. We assume that $\phi$ is a diffeomorphism onto its image.
Thus, $\FF$ has components $\FF =
(\FF_\phi, \FF_U, \FF_h)$, corresponding to the components of $B_2$, given by
\begin{subequations}\label{eq:Fdef}
\begin{align}
\FF_\phi &= \left ( \nabla_A(e^{\bar{U}} \bar{\sigma}_j{}^A) - e^{\bU}
[\bar{\epsilon} - \frac{\bar{\sigma}_l{}^l}{\bar{n}}]\,\overline{\partial_i
U} ,\quad \tr_{\partial\Bo} (\bar{\sigma}_i{}^A) n_A \right ) \\ \FF_U &=
\overline{\Delta_h U} - 4 \pi G e^{\bar{U}}(\bar{n} \bar{\epsilon} -
\bar{\sigma}_l{}^l) \chi_{\Bo} \\ \FF_h &= -\frac{1}{2}\overline{\Delta_h
h_{ij}} + Q_{ij}(\overline{h},\overline{\partial h}) - 2 \overline{\nabla_i
U} \overline{\nabla_j U} + 8 \pi G e^{\bar{U}} (\bar{\sigma}_{ij} -
\overline{h_{ij}}\,\bar{\sigma}_l{}^l)\chi_{\mathcal{B}}
\end{align}
\end{subequations}
Recall from the discussion in section \ref{sec:material} that the extension
$\hphi$ is needed in the definition of $\FF$.  The proof of the following
Lemma is a straightforward construction involving the use of the extension
operator $E$ and a cutoff function, and is left to the reader.
\begin{lemma} \label{lem:phiextend} Fix some $X_0 \in \Bo$.
There are constants $\mu > 0, R$ such that for each $\phi: \Bo \to
  \Re^3_{\Sp}$, $||\phi - \ii||_{W^{2,p}_{\Bo}} \leq \mu$, there is an
  extension $\hphi: \Re^3_{\Bo} \to \Re^3_{\Sp}$, which depends real
  analytically on $\phi$. The extension $\hphi$ can be chosen such that the
  map $\phi \mapsto \hphi$ is given by a continuous linear map from
  $W^{2,p}(\Bo)$ to $W^{2,p}(B_R(X_0))$, $\hphi(X) = \ii(X)$ for $X \in
  \Re^3_{\Bo} \setminus B_R(X_0)$, and $\hphi^{-1} \in
  W^{2,p}_{\loc}(\Re^3_{\Sp})$.
\end{lemma}
The equation to be solved is $\FF(G,Z)=0$.  The material form of the
reference state is given by
$$ Z_0 = (\ii, 0 , \hat \delta_{ij} \circ \ii) \in B_1 .
$$ The map $\FF$ defined by (\ref{eq:Fdef}) is easily verified to satisfy
$\FF(0,Z_0) = 0$ and to map $B_1 \to B_2$ locally near the reference state
$Z_0$.

\subsection{Differentiability of $\FF$}
In order to apply the implicit function theorem, we must verify that $\FF$ is
$C^1$ as a map $\Re \times B_1 \to B_2$, in the arguments $(G,Z)$, $Z = (\phi,\bU,
\overline{h_{ij}})$,  near $(0,Z_0)$. In fact, $\FF$ is real analytic if the
stored energy function $\eps$ is real analytic) near $Z_0$. It is clear the
dependence on $G$ is smooth. We will freely make use of the standard fact
that if $f$ is a smooth function, and $u \in W^{1,p}(\Bo)$, $p > 3$, then $u
\mapsto f(u)$ is a smooth mapping $W^{1,p}(\Bo) \to W^{1,p}(\Bo)$, as well as
the corresponding statement which holds for weighted Sobolev spaces. The map
$u \mapsto f(u)$ is sometimes called a Nemytskii operator.
We consider the dependence on $Z$ for each term separately.

It is straightforward to see that $\psi^A{}_i$ depends smoothly on $\phi$.
This means that in view of (\ref{Piola}) and the smoothness of the stored
energy function,  for $\FF_\phi$, we note that $\bar \sigma_i{}^A$ depends
smoothly on $Z$. Expanding the definition of $\FF_\phi$ it is clear this
depends smoothly on $Z$.

Next consider $\FF_U$. Note that $\overline{\Delta_h U}$ is the covariant
Laplacian in the metric $\bar H_{AB}$, where $\bar H_{AB}$ is the inverse of
$\bar H^{AB}$ given by (\ref{Hnew}), and thus $\overline{\Delta_h U}$ depends
smoothly on $Z$.  Using the fact that $\bar\sigma_i{}^j = \bar n \bar
\sigma_i{}^A \phi^j{}_{,A}$, the lower order terms in $\FF_U$ are seen to be
smooth in $Z$

Finally, we consider $\FF_h$. The discussion above of $\overline{\Delta_h U}$
applies also to $\overline{\Delta_h h_{ij}}$. The quadratic term
$Q_{ij}(\overline{h},\overline{\partial h})$ is evaluated by replacing each
occurrence of $h_{ij}$ by $\overline{h_{ij}}$ and $\partial_k h_{ij}$ by
$$ \psi^A{}_k \partial_A \overline{h_{ij}}
$$ In view of the regularity assumptions we find that
$Q_{ij}(\overline{h},\overline{\partial h})$ is smooth in $Z$. It is
straighforward to analyze the remaining terms in $\FF_h$ in the same manner.
We have now proved
\begin{lemma} The map $\FF: B_1 \to B_2$ is $C^1$ near $(0, Z_0)$.
\end{lemma}

\subsection{The Frechet derivative $D_2\FF(0,Z_0)$}
Next we
calculate the Frechet derivative $D_2 \FF(0,Z_0)$ and consider its
properties. As we shall see, this derivative is not an isomorphism $T_{Z_0}
B_1 \to B_2$, and therefore it will be necessary to modify the system of
equations by applying a projection before applying the implicit function
theorem.

We will denote by  $\zeta, v, k$ the infinitesimal variations of the fields
$\phi, \bU, \overline{h_{ij}}$.  Using the notation of section
\ref{sec:constitute}, let
$$\oS^{\ A\ B}_{i\ j} (X) = \delta^{AE}\delta^{BF}\delta^C{}_i \delta^D{}_j
\mathring{L}_{CEDF}(X) .
$$ The tensor $\oS^{\ A\ B}_{i\ j}$ corresponds to the tensor
$A_i{}^j{}_k{}^l$ of section \ref{sec:linelast}.  A calculation shows
\begin{align*}
D_\phi \bar\sigma_i^{\ A}(0, Z_0) . \zeta  &=  \oS^{\ A\ B}_{i\ j} \partial_B
\zeta^j \\ D_{\bU} \bar\sigma_i^{\ A}(0, Z_0) . v &= 2 \oS_i{}^A{}_j{}^B
\delta^j_B  v \\ D_{\bar h} \bar\sigma_i^{\ A}(0, Z_0) . k &=
\oS_i{}^A{}_j{}^B \delta^{lm} \delta_B^n k_{mn}
\end{align*}
Let $\xi^i$ be a Killing field in the reference metric $\hat\delta_{ij}$, in
Euclidean coordinates $\xi^i (x) = \alpha^i + \beta^i_{\ j} x^j$, with
$\alpha^i, \beta^i_{\ j}$ constants, $\beta^i_{\ j} = - \beta^j_{\ i}$, and
define $\xi^i(X) =\xi^i(\ii(X))$.  Then we have $\partial_A \xi^i =
\beta^i_{\ k} \delta^k_A$, and hence due to the antisymmetry of $\beta^i{}_k$,
\begin{equation}\label{eq:vanish}
\partial_A \xi^i \oS_i{}^A{}_j{}^B \equiv 0
\end{equation}
Let now $\delta \bar \sigma_i^{\ A}$ denote any combination of the Frechet
derivatives of $\bar \sigma_i^{\ A}$, evaluated at $(0, Z_0)$.  Assuming we
use a coordinate system $X^A$ where $V_{123} = 1$, we have due to
(\ref{eq:vanish}), and Stokes' theorem, the important relation
$$ 0 = \int_{\Bo} \xi^i \partial_A (\delta \bar \sigma_i^{\ A}) -
\int_{\partial\Bo} \xi^i (\delta \bar \sigma_i^{\ A}) n_A,
$$ where $n^A$ is the outward normal.  This can be interpreted as saying that
due to the natural boundary conditions, the linearized elasticity operator is
automatically equilibrated at the reference configuration $(0, Z_0)$.  It
follows from the constitutive conditions stated in section
\ref{sec:constitute}, that $ \zeta \mapsto D_\phi \FF_\phi (0, Z_0)\zeta$ is
elliptic, cf. the discussion in section \ref{sec:linelast}.  Therefore,
the operator
$$ D_\phi \FF_\phi (0, Z_0) : W^{2,p}(\Bo) \to [ L^p(\Bo) \times
B^{1-1/p,p}(\partial \Bo)]
$$ is Fredholm with kernel consisting of the Killing fields on $\Bo$, and
cokernel (in the sense of the natural $L^2$ pairing) consisting of the
Killing fields on $\ii(\Bo)$, as above.

The only nonzero contribution from the Frechet derivative of the first order
term
$$ - e^{\bU} [\bar{\epsilon} -
  \frac{\bar{\sigma}_l{}^l}{\bar{n}}]\,\overline{\partial_i U}
$$ in $\FF_\phi$ is
$$ D_{\bU} \left ( - e^{\bU} [\bar{\epsilon} -
\frac{\bar{\sigma}_l{}^l}{\bar{n}}]\,\overline{\partial_i U} \right )
\bigg{|}_{(0, Z_0)} . v = - \mathring{\eps} \partial_i v
$$ This is a lower order term which cannot affect the Fredholm property of
$D_2 \FF$, but it should be noted that it is not a priori equilibrated, and
therefore in general does not take values in the range of $D_\phi \FF_\phi$.
By the above discussion we have
$$ D_2 \FF_\phi(0,Z_0) = \left ( \partial_A (\delta \bar \sigma_i{}^A ) -
\mathring{\epsilon} \partial_i v, \quad \tr_{\partial\Bo}
(\delta\bar\sigma_i{}^A)n_A \right )
$$ Considering the other components of $D_2 \FF$, the only nonzero terms are
the diagonal entries $D_{\bU} \FF_U$ and $D_{\bar h} \FF_h$. These are given
by
$$ D_{\bU} \FF_U . v = \Delta v
$$ and
$$ (D_{\bar h} \FF_h . k)_{ij} = - \half \Delta k_{ij}
$$ where $\Delta = \delta^{AB} \partial_A \partial_B$ is the Laplacian in the
Euclidean background metric on $\Re^3_{\Bo}$. The operator $\Delta$ is an
isomorphism $W^{2,p}_{\delta} \to L^p_{\delta-2}$ for $\delta \in (-1,0)$,
cf. \cite{bartnik:mass}.

It follows from the above discussion that the Frechet derivative $D_2
\FF(0,Z_0)$ can be represented as the matrix of operators
$$
\begin{pmatrix} D_\phi \FF_\phi & D_U \FF_\phi & D_h \FF_\phi \\
0 & \Delta & 0 \\ 0 & 0 & -\half \Delta \end{pmatrix}
$$ (where the entries are evaluated at $(0, Z_0)$). In particular, the matrix
is upper triangular, and the diagonal entries are isomorphisms, with the
exception for $D_\phi \FF_\phi(0,Z_0)$ which is Fredholm with nontrivial
kernel and cokernel, cf. the discussion above.  The off diagonal terms are
bounded operators.  Therefore, if we compose $\FF$ with a projection which in
the first component maps onto the range of $D_\phi \FF_\phi(0,Z_0)$, and
restrict the domain of definition of $\FF$ to a subspace transverse to its
kernel, the resulting map will have Frechet derivative at $(0,Z_0)$ which is
an isomorphism, which will allow us to apply the implicit function
theorem. The projection which will be used is introduced in the next section.

\subsection{Projection} \label{sec:proj}
Introduce the projection operator $\BProj: B_2 \to B_2$, which acts as the
identity in the second and third components of $B_2$ and is defined in the
first component of $B_2$ as the unique projection along  the cokernel of
$D_\phi \FF_\phi(0,Z_0)$ onto the range of $D_\phi \FF_\phi (0,Z_0)$, which
leaves the boundary data in the first component of $B_2$
unchanged.  We use the the label $\Bo$ to indicate that $\BProj$ operates on
fields on the body and the extended body.  We shall later need to transport
the projection operator to fields on $\Re^3_{\Sp}$.

In order to give the explicit form of the action of  $\BProj$ in the first
component of $B_2$, let $(b_i, \tau_i)$ denote pairs of elements in
$L^p(\Bo) \times B^{1-1/p,p}(\partial \Bo)$. By a slight abuse of
notation, denote this operator too by $\BProj$.  We require that $\BProj
(b_i,\tau_i) = ({b'}_i , \tau_i)$, satisfying
\begin{equation}\label{eq:equilib}
\int_{\Bo} \xi^i {b'}_i = \int_{\partial \Bo} \xi^i \tau_i
\end{equation}
for all Killing fields $\xi^i$. Pairs $({b'}_i, \tau_i)$ satisfying this
condition will be called equilibrated.  Since $\BProj$ is a projection along
the cokernel, ${b'}_i$ must be of the form
$$ {b'}_i = b_i - \eta_i \chi_{\Bo}
$$ where $\eta_i = \alpha_i + \beta_{ij} X^j$, with $\alpha_i, \beta_{ij}$
constants satisfying $\beta_{ij} = - \beta_{ji}$.  We remark that the
bilinear pairing $(\xi, \eta) \mapsto \int_{\Bo} \xi^i \eta_i $ is
nondegenerate, it is simply the $L^2(\Bo)$ inner product on the space of
Killing fields. Further, the map
$$ \xi \mapsto \int_{\Bo} \xi^i b_i - \int_{\partial\Bo} \xi^i \tau_i
$$ defines a linear functional on the space of killing fields, for each pair
$(b_i, \tau_i)$. Therefore, there is a unique $\eta_i$ of the form given
above such that ${b'}_i = b_i - \eta_i \chi_{\Bo}$ satisfies
(\ref{eq:equilib}) for all Killing fields $\xi^i$.

\subsection{Existence of solutions to the projected system}
We are now in a position to apply the implicit function theorem to prove
\begin{prop} \label{prop:projected:implicit}
Let $\FF : B_1 \to B_2$ be map defined by (\ref{eq:Fdef}) and let $\BProj$ be
  defined as in section \ref{sec:proj}. Then, for sufficiently small values
  of Newton's constant $G$, there is a solution $Z = Z(G)$, where $Z = (\phi,
  \bU, \overline{h_{ij}})$, to the reduced, projected equation for
  self-gravitating elastostatics given by
\begin{equation}\label{eq:redproj}
\BProj \FF(G, Z) = 0 .
\end{equation}
In particular, for any $\eps > 0$, there is a $G  > 0$, such that $Z(G)$
satisfies the inequality
\begin{equation}\label{eq:Gbarsmall}
|| \phi - \ii ||_{W^{2,p}(\Bo)} + ||\overline{h_{ij}} - \delta_{ij}
||_{W^{2,p}_\delta} + ||\bU||_{W^{2,p}_\delta} < \eps .
\end{equation}

\end{prop}
\begin{proof}
Let $\YY$ denote the range of $\BProj$ and let $\XX$ be the subspace of
$T_{Z_0} B_1$, such that $(\phi - \ii)^i(X_0) = 0$ and $\delta^C{}_i
\delta_{C[A} \partial_{B]}(\phi - \ii)^i (X_0) = 0$ 
%
%
holds at some point $X_0 \in \Bo$.
We have already shown that $\FF: \Re \times B_1 \to B_2$ is $C^1$ and it
follows from the definition of $\BProj$ that $\BProj \FF: \Re \times \XX \to
\YY$ is $C^1$.  Since $\FF(0,Z_0)=0$, the Frechet derivative of $\BProj
\FF(G,Z)$ with respect to $Z$, evaluated at $(0,Z_0)$ is $\BProj D_2
\FF(0,Z_0)$ which we will denote by $A$.
It is clear from the discussion above that $A$ is Fredholm with trivial
cokernel. Therefore all that remains to be checked is that $\ker A$ is
trivial. To see this, note that $\ker D_\phi \FF_\phi$ consists of Killing
fields, by the discussion in section \ref{sec:linelast}. Prescribing the
value and antisymmetrized derivative of a Killing field at one point determines
the Killing field in all of $\Bo$.
This implies that $A : \XX \to \YY$ has trivial kernel, and it is
therefore an isomorphism. Thus, the implicit function theorem for Banach
spaces \cite{lang} applies to prove the existence of solutions to the
equation $\BProj \FF(G,Z) = 0$, for small values of $G$. Since $Z(G)$ depends
continuously on $G$ in $B_2$, the inequality (\ref{eq:Gbarsmall}) follows.
\end{proof}
\begin{remark} \label{rem:connected}
Up to this stage, the connectedness of the body $\Bo$ has not
  played at role in our arguments. In particular, in the case of a
body with $N > 1$ connected components, there is a collection of $N$ elastic
fields $\phi$ and an extension common to all of them. The equilibration
argument of section \ref{sec:bianchi} below fails, however, for the simple
reason that we have one equation, namely (\ref{ultimate}), but $N$ elastic
fields.
\end{remark}
In the next section, we study the solutions provided by Proposition
\ref{prop:projected:implicit}  and prove that they in fact
represent solutions of the full system (\ref{eq:Euler}).

\section{Equilibration} \label{sec:bianchi}
Let $(\phi, \bU, \overline{h_{ij}}) \in B_1$ be a solution to the reduced,
  projected system (\ref{eq:redproj}) as in Proposition
  \ref{prop:projected:implicit}, and
  let $\widehat{\phi}$ be the extension of $\phi$ provided by Lemma
  \ref{lem:phiextend}. Define $U$, $h_{ij}$ by
$U = \bU \circ \widehat{\phi}^{-1}$ and $h_{ij} = \overline{h_{ij}} \circ
  \widehat{\phi}^{-1}$.
Then $U$ and  $h_{ij}$ are elements of
  $W^{2,p}_\delta(\Re^3_{\Sp})$ and $E^{2,p}_\delta(\Re^3_{\Sp})$,
  respectively.  Further, we set $f = \widehat{\phi}^{-1}$. Then $f
  \in W^{2,p}_{\loc}$ and $f = \ii^{-1}$ outside a large ball.

Let $\hphi$ be the extension of $\phi$ provided by Lemma \ref{lem:phiextend},
and let $U, h_{ij}$ be defined by $U = \bU \circ \hphi^{-1}$, $h_{ij} =
\overline{h_{ij}} \circ \hphi^{-1}$. The following corollary to Proposition
\ref{prop:projected:implicit} is immediate  in view of the composition properties of
Sobolev functions.
\begin{cor} \label{cor:Gsmall}
For any $\eps > 0$, there is a $G > 0$ such
that the inequality
\begin{equation}\label{eq:Gsmall}
|| \phi - \ii ||_{W^{2,p}(\Bo)} + ||h_{ij} - \delta_{ij}
   ||_{W^{2,p}_\delta} + ||U||_{W^{2,p}_\delta} < \eps .
\end{equation}
holds.
\end{cor}

\subsection{Eulerian form of the projected system} \label{sec:eulerian}
Next we introduce the projection operator in space which corresponds to
$\BProj$. From this point on, we are only interested in the
action of $\BProj$ in the first component of $B_1$.
We will be dealing with solutions to the projected system $\BProj \FF = 0$,
in particular this means that the boundary condition $\bar\sigma_i{}^A n_A
\big{|}_{\partial \Bo} = 0$ will be satisfied, due to the nature of the
projection operator defined in section \ref{sec:proj}.
Therefore it is convenient to write $\BProj b = \BProj(b,0)$.

Recall the change of variables formula
$$
\int_{\Bo} u (X) dV(X) = \int_{f^{-1}(\Bo)} u \circ f (x) n(x) d\mu_h(x)
$$
which results from the definition of $n$ in equation (\ref{n}). This implies
$$
\int_{\Bo} \xi^i(X) b_i(X) dV(X) = \int_{f^{-1}(\Bo)} \xi^i \circ f (x)
n(x) b_i \circ f(x)
d\mu_h
$$
Letting $b_i = \nabla_A(e^{\bar{U}} \bar{\sigma}_j{}^A)-
e^{\bar{U}}( \bar{\epsilon} -
\frac{\bar{\sigma}_l{}^l}{\bar{n}})\,\overline{\partial_j
  U}$, we have, using the properties of the Piola transform,
$$
n b_i \circ f = \nabla_j (e^U \sigma_i{}^j) - e^U ( n \eps -
\sigma_l{}^l) \partial_i U
$$
Based on this, we define the space version $\fProj$ of the projection
$\BProj$ by
$$
\fProj (n \cdot (b \circ f)) = n  ( \BProj b ) \circ f
$$
$\BProj$ is a projection by construction, and it follows from the definition
of $\fProj$ that it also is a projection.
Due to this definition, the equation
$\BProj \FF_\phi  = 0$ implies $\fProj(n \FF_\phi \circ f) = 0$. This relation
is written more explicitely as equation (\ref{projelast1})
below. Summarizing, the  triple $(\phi,U,h_{ij})$ we have
constructed satisfies the following set of equations,
\begin{subequations} \label{eq:projredstat}
\begin{align}
\fProj \left(\nabla_j(e^{U} \sigma_i{}^j)-
e^{U}( n \eps - \sigma_l{}^l ) \partial_i U \right) & =0
\quad \text{ \rm in } f^{-1} (\Bo) , \label{projelast1} \\
\sigma_i{}^j n_j|_{\partial
f^{-1}(\Bo)}&=0 \quad \text{\rm in } \partial f^{-1}(\Bo) \label{bound1} \\
\Delta_h U &= 4 \pi G e^U(n \epsilon - \sigma_l{}^l)
\chi_{f^{-1}(\Bo)} \quad \text{\rm in} \Re^3_{\Sp} \label{potential1} \\
G_{ij} - (\nabla_{(i} V_{j)} - \frac{1}{2} h_{ij} \nabla_l V^l)  &=
8 \pi G ( \Theta_{ij} - e^U
\sigma_{ij}\,\chi_{f^{-1}(\Bo)})\quad \text{\rm in}\;\Re^3_\Sp \label{reduced1}
\end{align}
\end{subequations}
Recall that equation (\ref{reduced1}) is equivalent to the space version of
equation (\ref{barreduced}),
\begin{equation}\label{Ricreduced1}
-\half \Delta_h h_{ij} +
Q_{ij}(h,\partial h) = 2 \nabla_iU
\nabla_j U - 8 \pi G e^{U} (\sigma_{ij} -
h_{ij} \sigma_l{}^l) \chi_{f^{-1} (\Bo)}.
\end{equation}
Since the body is bounded, the right hand sides of equation
(\ref{potential1}) and the contribution from the stress in equation
(\ref{Ricreduced1}) have compact support, and hence $U, h_{ij}$ satisfy the
reduced vacuum static Einstein equations near spatial infinity. This implies that we
can get more information about their asymptotic behavior than is a priori
given from the implicit function theorem argument.
\begin{lemma} \label{lem:asympt-stat}
$U, h_{ij}$ are of the form
\begin{align}
U &= \frac{m_U}{r} + U_{(2)} \label{eq:Uform} \\
h_{ij} &= \delta_{ij} + \frac{\gamma_{ij}}{r}  + h_{(2)\, ij}
\label{eq:hform}
\end{align}
for constants $m_U, \gamma_{ij}$, with $h_{(2)\, ij}, U_{(2)} \in W^{2,p}_{2\delta}$.
Further, for sufficiently small
$G$,
we have the estimates
\begin{align}
||U_{(2)}||_{W^{2,p}_{2\delta}} + |m_U| &\leq C ( ||h_{ij} - \delta_{ij}
  ||_{W^{2,p}_{\delta}}
+ ||\phi - \ii||_{W^{2,p}(\Bo)}
)
   \label{eq:Ubound}  \\
||h_{(2)\ ij} ||_{W^{2,p}_{2\delta}} + ||\gamma|| &\leq C ( ||h_{ij} - \delta_{ij}
  ||_{W^{2,p}_{\delta}}
+ ||\phi - \ii||_{W^{2,p}(\Bo)}
)
\label{eq:hbound}
\end{align}
where $||\gamma|| = ( \sum_{i,j} \gamma_{ij}^2 )^{1/2}$.
\end{lemma}
\begin{remark}
If $h_{ij}, U$ were solutions of the full static vacuum Einstein equations
near infinity, with the same falloff conditions, one could conclude 
that the
$O(1/r)$ term in $h_{ij}$ vanishes, following the argument in
\cite{kennefick:omurchadha}, see also \cite{beig:simon:1981}.
However, in our situation $U, h_{ij}$
satisfies the reduced static vacuum Einstein equations near infinity and this
argument does not apply directly.
\end{remark}
\begin{proof}
By construction, we have $h_{ij} - \delta_{ij} \in W^{2,p}_{\delta}$ and $U
\in W^{2,p}_{\delta}$, with $\delta \in (-1,-1/2)$.
This, together with equation (\ref{potential1})
implies using the product estimates (\ref{eq:prodweighted}) to expand
$\Delta_h U$, that
$$
\Delta_e U  = f \in L^p_{2\delta-2}
$$
with
\begin{align*}
||f||_{L^p_{2\delta-2}} &\leq C ( ||h_{ij} - \delta_{ij} ||_{W^{2,p}_{\delta}}
 + ||n \epsilon - \sigma_l{}^l||_{L^p(f^{-1}(\Bo)} ) \\
&\leq  C ( ||h_{ij} - \delta_{ij} ||_{W^{2,p}_{\delta}}
+ ||\phi - \ii||_{W^{2,p}(\Bo)}
 ) ,
\end{align*}
where in the last step we estimated $n\eps - \sigma_l{}^l$ in terms of
$h_{ij}$, $\phi - \ii$.

We follow the proof of \cite[Theorem 17]{bartnik:mass}, see also the
discussion in section \ref{sec:weighted}.
Let $u$ satisfy
$$
\Delta_e u = f \in L^p_{2\delta -2}
$$
By the isomorphism property of $\Delta_e : W^{2,p}_{\delta} \to
L^p_{\delta-2}$, and the inclusion $L^p_{2\delta-2} \subset L^p_{\delta-2}$,
we have
$$
||u||_{W^{2,p}_{\delta}} \leq C ||f ||_{L^p_{2\delta-2}}
$$
Fix some large radius $R$, and a point $x_0 \in \Re^3$.
Let $B_R = B_R(x_0)$, and let $E_R = \Re^3 - B_R$.
By assumption, $\delta \in (-1,-1/2)$, so that $-2 < 2\delta < -1$. In
particular $2\delta$ is nonexceptional, and hence $\Delta_e:
W^{2,p}_{2\delta} \to L^p_{2\delta -2}$ is Fredholm.
Hence there is a $v \in W^{2,p}_{2\delta}$ such that
$$
\Delta_e (u - v) = 0 \quad \text{\rm in } E_R
$$
It follows that $u-v$ satisfies
$$
u-v = \frac{m}{r} + O(1/r^2) \quad \text{\rm in } E_R
$$
for some constant $m$. Here the term $O(1/r^2)$ comes from the expansion of
harmonic functions, and is in $W^{2,p}_{2\delta}$.
Let $\zeta$ be a cutoff function such that $\zeta = 1$ in $E_{2R}$, $\zeta = 0$
in $B_R(x_0)$ and with
$|\partial \zeta| \leq C/R$. Set
$$
w = u - \frac{m}{r} \zeta \in W^{2,p}_{2\delta}
$$
Then we have $w = u$ in $B_R(x_0)$. Further, we have
$$
\Delta_e w = f - \Delta_e (\frac{m}{r}\zeta)
$$
where $\Delta_e (\frac{m}{r}\zeta)$ has compact support and we can estimate
$$
||\Delta_e(\frac{m}{r} \zeta)||_{L^2_{2\delta -2}} \leq C|m|
$$
Since
$$
u = \Gamma \star f
$$
where $\Gamma = 1/r$ is the fundamental solution of $\Delta_e$ we have
$$
|m| \leq C ||f||_{L^1(\Re^3)} \leq C ||f||_{L^p_{2\delta-2}}
$$
Here we used the inclusion $L^p_{2\delta-2} \subset L^1_{2\delta-2} \subset
L^1_{-3}$, and the fact that $L^1 = L^1_{-3}$ in 3 dimensions.
Thus we have
$$
||\Delta_e w||_{L^p_{2\delta-2}} \leq C ||f ||_{L^p_{2\delta-2}}
$$
Using the fact that $\ker \Delta_e$ is trivial in $W^{2,p}_{2\delta}$
together with the Fredholm property, we now have
$$
||w||_{W^{2,p}_{2\delta}}
\leq
C ||f||_{L^p_{2\delta-2}}
$$
This argument proves (\ref{eq:Uform}) and (\ref{eq:Ubound}). Applying the
same argument to $h_{ij}$ shows
$$
h_{ij} = \delta_{ij} + \frac{\gamma_{ij}}{r} + h_{(2)\, ij}
$$
with $\gamma_{ij}$ constant and $h_{(2)\,ij} \in
W^{2,p}_{2\delta}$, and gives an estimate for $h_{(2)\, ij}, \gamma_{ij}$ in
terms of $h_{ij} - \delta_{ij}$, $U$, and $\phi - \ii$. This gives
(\ref{eq:hbound}), since we have
an estimate for $U$ in terms of $h_{ij} - \delta_{ij}$, $\phi - \ii$ from
(\ref{eq:Ubound}).
\end{proof}

\subsection{The tension field}
The next three lemmas state the properties of the tension field $V^k$,
defined by (\ref{eq:Vdef}), which we
shall need. We may assume $\hGamma^i_{jk} = 0$.
Let the operator $L$ be defined by
$$
LV^k = \Delta_h V^k + R^k{}_j V^j
$$
where $\Delta_h$, $R_{ij}$
are the covariant Laplacian and the Ricci tensor of
$h_{ij}$, respectively.
\begin{lemma} \label{lem:Vform}
Assume we are in a Cartesian coordinate system so that
$\hat{\delta}_{ij} = \delta_{ij}$ and $\hGamma_{ij}^k = 0$. Let $r = |x|$ and
  set $\theta^i = x^i/r$.
Let $h_{ij}$, $\gamma_{ij}$ be as in (\ref{eq:hform}).
Suppose $L V \in L^p_{\delta-3}$, where $V$ is the tension field of $h$.
Then if $G$ is sufficiently small, $V \in
W^{2,p}_{\delta-1}$ and the estimate
\begin{equation} \label{eq:Lest}
|| V ||_{W^{2,p}_{\delta-1}} \leq C
|| LV||_{L^p_{\delta-3}}
\end{equation}
holds. Further,
$V$ is of the form
\begin{equation}\label{eq:Vform}
V^k = -\frac{1}{2r^2} \left ( \gamma^k{}_j \theta^j - \gamma^m{}_m \theta^k
  \right ) + V^k_{(3)}
\end{equation}
where
$V^k_{(3)} \in W^{2,p}_{2\delta -1}$, and $\gamma^i{}_j =
\delta^{ik}\gamma_{kj}$.
\end{lemma}
\begin{proof}
The operator $L$ is an elliptic system, which is of diagonal type. Therefore
the results discussed in section \ref{sec:weighted} apply to $L$. In
particular, $L$ is asymptotic to $\Delta$ with rate $\delta$. Thus,
$L : W^{2,p}_{\delta-1} \to L^p_{\delta-3}$
is Fredholm. It follows from the
definition of $V$ and the assumptions on $h$ that $V \in
W^{1,p}_{\delta-1}$. If $LV \in L^p_{\delta-3}$, then elliptic regularity
gives $V \in W^{2,p}_{\delta-1}$. For small data $L$ has trivial kernel
in this range of
spaces, since $L$ is then a small perturbation of $\Delta_e$ which has
trivial kernel in $W^{2,p}_{\tau}$, $\tau < 0$. By Corollary
\ref{cor:Gsmall}, for $G$ sufficiently small, the required condition on $L$
will hold. Therefore, if $G$ is sufficiently small, an estimate of the form
(\ref{eq:Lest}) holds.

Recall from Lemma \ref{lem:asympt-stat} that
$$
h_{ij} = \delta_{ij} + \frac{\gamma_{ij}}{r} + h_{(2)\, ij}
$$
with $h_{(2)\, ij} \in W^{2,p}_{2\delta}$.
Let
\begin{equation}\label{eq:Hdef}
H^k = -\frac{1}{2r^2} \left ( \gamma^k{}_j \theta^j - \gamma^m{}_m \theta^k
  \right ) .
\end{equation}
Under our assumptions,
$\hGamma_{ij}^k = 0$. From the definition of $V^k$, a calculation shows
$$
V^k - H^k \in W^{1,p}_{2\delta-1},
$$
in particular $V^k - H^k = o(1/r^{2\delta-1})$.
The operator $L$ is asymptotic to $\Delta_e$ with rate $\delta$, see section
\ref{sec:weighted} for this notion.
It follows by an argument along the lines of the proof of Lemma
\ref{lem:asympt-stat} that
$$
V^k = H^k + V^k_{(3)}
$$
with $V^k_{(3)} \in W^{2,p}_{2\delta-1}$.
This completes the proof of the Lemma.
\end{proof}

Let $\xi$ be a Euclidean Killing field of the form
$\xi^i(X) = \xi^i (\ii(X))$ with $\xi^i(x) = \alpha^i +
\beta^i{}_j x^j$ for $\alpha^i, \beta^i{}_j$ constants such that
$\beta^i{}_j = - \beta^j{}_i$.

\begin{lemma} \label{lem:part-int}
Let $V$ be as in (\ref{eq:Vform}). Then the identity
\begin{equation}\label{eq:part-int}
\int \xi^i \circ f LV_i d\mu_h = \int L(\xi^i \circ f) V_i d\mu_h
\end{equation}
holds.
\end{lemma}
\begin{proof}
First note that both sides of equation (\ref{eq:part-int}) are well defined.
The integrand on the left hand side has compact support by equation
(\ref{ultimate}). Recall
that
$f$ equals
the identity map outside of a compact set. We have
$$
L(\xi\circ f) \in L^p_{\delta-1}
$$
and
$$
V_i = O(1/r^2)
$$
Therefore the integrand in the right hand side is an element of
$L^p_{\delta-3} \subset L^1$.
In order to justify the partial integration, let $B_R = \{ |x| \leq R\}$ and
$S_R = \{ |x|=R\}$. Then (\ref{eq:part-int}) is equivalent to
$$
\lim_{R \to \infty} \int_{B_R}  \xi^i \circ f LV_i d\mu_h =
\lim_{R \to \infty} \int_{B_R}  L(\xi^i \circ f) V_i d\mu_h .
$$
Gauss law applied to the integral over $B_R$ gives
\begin{multline}\label{eq:LVboundary}
\int_{B_R} \xi^i \circ f
LV_i d\mu_h = \int_{B_R} L(\xi^i \circ f) V_i d\mu_h \\
+
\int_{S_R} \xi^i\circ f \nabla_k V^i n^k dA_h
- \int_{S_R} \nabla_k (\xi^i \circ f) n^k V_i dA_h
\end{multline}
Let $\theta^k = x^k/r$. With
the asymptotic behavior of $h_{ij}$ from Lemma \ref{lem:asympt-stat} we
have $n^k = \theta^k + O(1/r)$ and
$\Gamma^i_{jk} = O(1/r^2)$. Further, the area element induced on $S_R$ from
$h_{ij}$ differs from the standard area element of $S_R$ by a term of order
$O(1/R)$. Recall that by Lemma \ref{lem:Vform},
$V^k = H^k + V^k_{(3)}$ with $H^k$ given by (\ref{eq:Hdef}) and $V^k_{(3)}
\in W^{2,p}_{2\delta-1}$.
Since we are interested in the integral over
$S_R$ in the limit $R \to \infty$, this shows that the only important terms in
the  boundary integrals in (\ref{eq:LVboundary}) are
\begin{equation}\label{eq:boundasy}
\int_{S_R} \beta^k{}_i x^i \partial_l H_k \theta^l dA_R
- \int_{S_R} \beta^i{}_k H^i \theta^k dA_{R}
\end{equation}
We consider the first of these terms.
Thus let $Z = \beta^k{}_i x^i \partial_l H_k \theta^l$,
the integrand in the first boundary integral above.
At this point, we will be doing all calculations in the background metric
$\delta_{ij}$, and therefore it is convenient to lower all indices using
$\delta_{ij}$ and sum over repeated indices. A calculation shows
$$
\partial_l H_k = - \frac{1}{2r^3} \left [ 2 \gamma_{lk} - \gamma_{mm}
  \delta_{lk} - 3 ( 2 \gamma_{jk} \theta_l \theta_j - \gamma_{mm} \theta_l
  \theta_k ) \right ]
$$
This gives using $\theta_k \theta_k = 1$,
$$
Z = \beta_{ki} x_i \partial_l H_k \theta_l = - \frac{1}{2r^2} \beta_{ki}
\theta_i ( - 4 \gamma_{jk} \theta_j + 2 \gamma_{mm} \theta_k )
$$
Note
that the second term in the right hand side vanishes due to the fact that
$\beta_{ki}
= - \beta_{ik}$, but not the first. To show this vanishes we proceed as
follows.
We have $dA_R = R^2 dA$ where $dA$ is the area element of the unit sphere
$S$, so we can write $\int_{S_R} Z dA_R = R^2 \int_{S} Z
dA$. Recall the identity
$$
\int_{S} \theta_k \theta_l dA = \frac{4\pi}{3} \delta_{kl} \, .
$$
This gives
$$
\int_{S_R} Z dA_R = -\frac{2\pi}{3} \beta_{ki} ( - 4 \gamma_{jk} \delta_{ij} + 2
\gamma_{mm} \delta_{ik} ) = 0
$$
due to the fact that $\beta_{ki} = - \beta_{ik}$.
The integrand in the second term in (\ref{eq:boundasy})
is
$$
-\frac{1}{2r^2} \beta_{ik} (\gamma_{ij}\theta_j - \gamma_{mm} \theta_i)
\theta_k
$$
which can be handled using the same method.
It follows that
$$
\lim_{R \to \infty} \left ( \int_{S_R}  \xi^i\circ f \nabla_k V^i n^k dA_h -
\int_{S_R} \nabla_k (\xi^i \circ f) n^k V_i dA_h \right )
= 0
$$
which completes the proof of the Lemma.
\end{proof}

\begin{lemma} \label{lem:xiLV-est}
For sufficiently small $G$, there is a constant $C$ such that the
inequality
$$
|\int (\xi^i \circ f) L V_i d\mu_h | \leq C\left ( ||h_{ij} - \delta_{ij}
  ||_{W^{2,p}_{\delta}}
+ ||\phi - \ii||_{W^{2,p}(\Bo)} \right ) |\xi| ||V||_{W^{2,p}_{\delta-1}}
$$
holds, where for $\xi^i = \alpha_i + \beta^i{}_j x^j$, we write $|\xi| =
|\alpha| + |\beta|$.
\end{lemma}
\begin{proof}
Using Lemma \ref{lem:part-int} we consider
$$
\int L (\xi^i \circ f) V_i d\mu_h
$$
We need to estimate $||L(\xi^i \circ f) V_i ||_{L^1}$. We expand out $L(\xi
\circ f)$, dropping the reference to $f$ for brevity, and writing $\star$ to
denote a general contraction
$$
L\xi = h \star \partial^2 \xi + \Gamma \star \partial \xi + \partial \Gamma
\star \xi + R \star \xi .
$$
Using the form of $h_{ij}$ given in equation (\ref{eq:hform}), we have
\begin{align*}
\Gamma &=   O(1/r^2) + v, \quad \text{\rm with } v \in L^p_{2\delta-1} \\
\partial\Gamma &= O(1/r^3) + z, \quad \text{\rm with } z \in L^p_{2\delta-2}
\\
R &= O(1/r^3) + w, \quad \text{\rm with } w \in L^p_{2\delta-2}
\end{align*}
Further, using the estimates from Lemma \ref{lem:asympt-stat}, the
coefficients in the $O(1/r^2)$ and $O(1/r^3)$ terms can be estimated in terms
of $|\gamma|$, where $\gamma_{ij}$ is the constant metric in
(\ref{eq:hform}), and the lower order terms can be estimated in terms of
$h_{(2)\, ij}$. The term involving $\partial^2 \xi$ has compact support and
can be estimated in terms
of $\phi - \ii$. Multiplying $L \xi \circ f$ by $V$, using $\xi = O(r)$ and
$V \in W^{2,p}_{\delta-1}$, we find the result gives
terms in $L^p_{\delta-3}$ and $L^p_{3\delta-2}$ coming from the $O(1/r^2),
O(1/r^3)$ terms and the Sobolev terms, respectively. Since $L^p_{3\delta-2}
\subset L^p_{\delta-3}$ and $L^p_{\delta-3} \subset L^1$, this means that
using Lemma \ref{lem:asympt-stat} we have
an estimate of the form
\begin{align*}
||L(\xi \circ f) V||_{L^1} &\leq C \left (|\gamma| + ||h_{(2)\,
  ij}||_{W^{2,p}_{2\delta}}
 + ||\phi - \ii||_{W^{2,p}_{\Bo}} \right ) |\xi | \, ||V||_{W^{2,p}_{\delta-1}} \\
&\leq C \left ( ||h_{ij} - \delta_{ij} ||_{W^{2,p}_{\delta}}
+ ||\phi - \ii||_{W^{2,p}(\Bo)} \right ) |\xi| \,
||V||_{W^{2,p}_{\delta-1}}
\end{align*}
This completes the proof.
\end{proof}
We now take, in the sense of distributions, the divergence of
equation (\ref{reduced1}), i.e. the equation
\begin{equation}\label{reduced1-late}
G_{ij} - (\nabla_{(i} V_{j)} - \frac{1}{2} h_{ij} \nabla_l V^l)  =
8 \pi G ( \Theta_{ij} - e^U
\sigma_{ij}\,\chi_{f^{-1}(\Bo)}) \,,
\end{equation}
which holds in $\Re^3_{\Sp}$.
For the first term on the left in equation (\ref{reduced1-late}) we
use the Bianchi identity from Lemma \ref{lem:bianchi} to conclude $\nabla^i
G_{ij} = 0$ in the sense of distributions. In order to take the divergence of
the second term on the left, note that it follows from
the definition of $V$ that $V \in W^{1,p}_{\delta -1}$.
The regularity assumptions on $h_{ij}$ imply that the identity
$$
\nabla^j ( \nabla_{(i}V_{j)} - \half h_{ij} \nabla_l V^l) = \half ( \Delta_h V_i +
R_{ij} V^j )
$$
holds in the sense of distributions.
For the first term on the right of (\ref{reduced1-late}), we use
equation (\ref{potential1}). In the second term on the right in
(\ref{reduced1-late}), using that the boundary condition
(\ref{bound1}) is satisfied, we apply Lemma \ref{lem:div} to get the
identity
$$
- \nabla^j ( \Theta_{ij} - e^U \sigma_{ij} \chi_{f^{-1}(\Bo)} ) =
\left[\nabla_j (e^U \sigma_i{}^j) -
e^U (n \epsilon - \sigma_l{}^l) \partial_i U \right]\;\chi_{f^{-1}(\Bo)}
$$
In view of the projected elastic equation (\ref{projelast1}), this
is equivalent to
\begin{multline*} 
- \nabla^j ( \Theta_{ij} - e^U \sigma_{ij} \chi_{f^{-1}(\Bo)} ) = \\
(\fId - \fProj) \left[\nabla_j (e^U \sigma_i{}^j) -
e^U (n \epsilon - \sigma_l{}^l) \partial_i U \right]\;\chi_{f^{-1}(\Bo)}
\end{multline*} 
where $\fId$ is the identity operator in the space $L^p(f^{-1}(\Bo))$.
Thus, defining the operator $L$ by
$$
LV_i = \Delta_h V_i + R_{ij} V^j ,
$$
we finally conclude that the equation
\begin{equation}\label{ultimate}
L V_i = 16 \pi G (\fId - \fProj) \left [ \nabla_j (e^U \sigma_i{}^j) -
e^U (n \epsilon - \sigma_l{}^l) \partial_i U \right ]
\chi_{f^{-1}(\mathcal{B})}
\end{equation}
holds in $\Re^3_{\Sp}$.

We observe that equation (\ref{ultimate}) implies that $LV$
is supported in $f^{-1} (\Bo)$. Further,
in view of the fact that $\fProj$ is a projection, we have
\begin{equation}\label{eq:PLV}
\fProj L V = 0 .
\end{equation}
Recall from section \ref{sec:analytical}
that $\BProj$ was defined in terms of
the Killing fields $\xi^i(X) = \xi^i (\ii(X))$ with $\xi^i(x) = \alpha^i +
\beta^i{}_j x^j$ for $\alpha^i, \beta^i{}_j$ constants such that
$\beta^i{}_j = - \beta^j{}_i$. It
follows from the definition of $\fProj$, that we have
$$
0 = \int_{\Re^3_{\Sp}} (\xi^i \circ f(x))  \fProj z_i \chi_{f^{-1}(\Bo)}
  d\mu_h
$$
for any $\xi^i$ of the above form and any $z_i \in L^p_{\loc}(\Re^3_{\Sp})$.
Now define the linear
mapping $\QQ : L^p_{\delta - 3}(\Re^3_{\Sp}) \to \Re^6$, by setting
$$
\QQ_{\kappa} (z_i) = \int_{\Re^3_{\Sp}} (\xi^i_{(\kappa)} \circ f )z_i d\mu_h , \quad
\kappa = 1,\dots,6
$$
where $\{ \xi_{(\kappa)} \}_{\kappa = 1}^6$ forms a basis for the space of Killing fields.
We have the estimate
\begin{lemma} \label{lem:Qbound}
There is a constant, depending on $f$ and $h_{ij}$ such that the inequality
\begin{equation}\label{eq:Qbound}
|| z  ||_{L^p(f^{-1}(\Bo))} \leq C ( ||\fProj
z||_{L^p(f^{-1}(\Bo))} + ||\QQ z \chi_{f^{-1}(\Bo)} ||_{\Re^6})
\end{equation}
holds.
\end{lemma}
\begin{proof}
From the definition of $\fProj$, we have
$$
z \chi_{f^{-1}(\Bo)} = \fProj z \chi_{f^{-1}(\Bo)} + n (\zeta \circ f)
\chi_{f^{-1}(\Bo)}
$$
where $\zeta$ is a Killing field, and from the definition of $\QQ$, we have
\begin{align*}
\QQ_{(\kappa)} z \chi_{f^{-1}(\Bo)} &=
\int_{\Re^3_{\Sp}} (\xi^i_{(\kappa)} \circ f)
n (\zeta \circ f) \chi_{f^{-1}(\Bo)} d\mu_h \\
&= \int_{\Bo} \xi^i_{(\kappa)} \zeta_i dV
\end{align*}
for a basis of Killing fields $\{\xi_{(\kappa)}\}_{\kappa = 1}^6$.
In view of the non-degeneracy of the $L^2$ pairing on the space of Killing
fields on $\Bo$, this means that there is a constant $C$ such that
$|| n (\zeta \circ f) ||_{L^p(f^{-1}(\Bo))} \leq C ||\QQ  n (\zeta \circ f)
\chi_{f^{-1}(\Bo)} ||_{\Re^6}$. The inequality (\ref{eq:Qbound}) now follows
after an application of the triangle inequality.
\end{proof}

\begin{prop}\label{prop:VQbound}
For sufficiently small $G$, there is a constant $C$ such that the estimate
\begin{equation} \label{eq:VQbound}
||V||_{W^{2,p}_{\delta-1}} \leq C ||\QQ LV||_{\Re^6}
\end{equation}
holds.
\end{prop}
\begin{proof}
Due to the boundedness of $\Bo$ and $f^{-1}(\Bo)$, there is a constant $C$
such that
$$
||z \chi_{f^{-1}(\Bo)} ||_{L^p_{\delta-3}} \leq C || z
||_{L^p(f^{-1}(\Bo))}
$$
Since $V$ solves (\ref{ultimate}), $LV$ is supported in $f^{-1}(\Bo)$ so the
above inequality applies to $LV$. Now making use of
(\ref{eq:PLV}), the estimate (\ref{eq:Qbound}), together with the estimate
(\ref{eq:Lest}) from Lemma \ref{lem:Vform}, we
find that for $h_{ij}$ sufficiently close to $\delta_{ij}$, i.e. for $G$
suffiently close to $0$, the estimate (\ref{eq:VQbound}) holds.
\end{proof}

%
%
%
%
\subsection{The main theorem}
We are now able to prove
\begin{thm} \label{thm:main}
For sufficiently small values of $G$, the solution to the reduced, projected,
system of equations for a static, elastic, self-gravitating body, equations
(\ref{eq:projredstat}) is a solution to the full system (\ref{eq:Euler})
of equations for a
static, elastic, self-gravitating body.
\end{thm}
\begin{proof}
Recall the definition of $\QQ$. We have
$$
\QQ_{(\kappa)} LV = \int (\xi^i_{(\kappa)} \circ f) L V_i d\mu_h
$$

%


For small data, i.e. for sufficiently
small $G$, $\xi$ is an approximate Killing field for $h_{ij}$.
Therefore by Lemma
\ref{lem:xiLV-est}, there is for sufficiently small $G$ a constant $C$ such
that
$$
||\QQ LV||_{\Re^6} \leq C\left ( ||h_{ij} - \delta_{ij} ||_{W^{2,p}_{\delta}}
+ ||\phi - \ii||_{W^{2,p}(\Bo)} \right )   ||V||_{W^{2,p}_{\delta-1}}
$$
By (\ref{eq:VQbound}), this implies that after possibly decreasing $G$, there
is a constant $C$ such that
$$
||V||_{W^{2,p}_{\delta-1}} \leq C\left ( ||h_{ij} - \delta_{ij} ||_{W^{2,p}_{\delta}}
+ ||\phi - \ii||_{W^{2,p}(\Bo)} \right )   ||V||_{W^{2,p}_{\delta-1}}
$$
Hence, by Corollary \ref{cor:Gsmall}, for
sufficiently small values of $G$, we have
$$
||V||_{W^{2,p}_{\delta-1}} \leq \half ||V||_{W^{2,p}_{\delta-1}}
$$
Thus, in fact $V = 0$.
Hence equation (\ref{reduced1})
implies the full Einstein equation (\ref{metric}).
Further, by
(\ref{ultimate}) we have that
$$
(\fId - \fProj) \left[\nabla_j (e^U \sigma_i{}^j) -
e^U (n \epsilon - \sigma_l{}^l) \partial_i U \right]\;\chi_{f^{-1}(\Bo)} = 0
$$
while by equation (\ref{projelast1}), 
$$
\fProj \left(\nabla_j (e^U \sigma_i{}^j) -
 e^{U}( n \eps - \sigma_l{}^l ) \partial_i U \right)  =0
\quad \text{ \rm in } f^{-1} (\Bo)
$$
It follows that equation (\ref{elast}) is satisfied, and hence that
the solution $(\phi, U, h_{ij})$ to the projected, reduced system, which was
constructed using the implicit function theorem, is a solution to the full
system (\ref{eq:Euler}).
\end{proof}

        

\appendix


%
\section{}\label{sec:bobbyapp}
The way the spatial metric $h_{ij}$ is treated in section
\ref{sec:analytical} may seem surprising at
first.  We start by reducing the Einstein equation
(\ref{metric}) using harmonic gauge, and then pull
back the components of $h_{ij}$ in that gauge. Geometrically it might seem
much more natural to start by pulling back the metric itself under
$\hat{\phi}$. Doing this would result in replacing
(\ref{metric})
by
$$
   \bar{G}_{AB} = 8 \pi G (\bar{\Theta}_{AB} - e^{-\bar{U}} \bar{\sigma}_{AB}
\chi_{\mathscr{B}})\hspace{0.5cm}\textrm{in}\hspace{0.2cm}
\mathbb{R}_{\mathscr{B}}^3,
$$
where $\bar{G}_{AB}$ is the Einstein tensor of $\bar{H}_{AB}$. Furthermore,
we have that
$$
\bar{\Theta}_{AB} = \frac{1}{8 \pi G}[D_A \bar{U} D_B \bar{U} - \frac{1}{2}
\bar{H}_{AB} \bar{H}^{CD} D_C \bar{U} D_D \bar{U}]
$$
and $\bar{\sigma}_{AB}$ is the extension of $\sigma_{AB}$. But then the
quantity $\bar{\sigma}_i{}^A$ in equation
(\ref{Piola})
depends on $\phi$ only
through $\widehat{\psi}^A{}_i$, since the remaining factors in
\begin{equation}\label{eq:1}
\bar{\sigma}_i^A = \widehat{\psi}^B{}_i J \bar{\sigma}_{BC} \bar{H}^{AC}
\end{equation}
depend only on $\bar{H}_{AB}$. Let us now look at the nature of Eq.
(\ref{elastmaterial}),
when $(\phi, \bar{U}, \bar{H}_{AB})$ are used as the basic variables rather
than $(\phi, \bar{U},\bar{h}_{ij})$. There are potentially two terms
depending on second derivatives of $\phi$. One comes from taking a derivative
of $\widehat{\psi}^A{}_i$ in equation (\ref{eq:1})
above. The other comes from writing the
Christoffel symbols of $h_{ij}$ entering the left hand side of
(\ref{elastmaterial})
in terms of
$\phi$. Remarkably, both these terms cancel!
For the linearization of the residual
of equation
(\ref{elastmaterial})
at $(\phi = {\bf{i}}, \bar{U}=0,\bar{H}_{AB}=\delta_{AB})$ the
only surviving term will involve first partial derivatives of $\bar
{H}_{AB}$, i.e., terms which break diffeomorphism invariance on
$\Re^3_{\Bo}$.  However note that the very presence of the a
priori given domain $\Bo$ (which of course is the reason why this
whole manoeuvre was performed) breaks diffeomorphism invariance. It is thus
no wonder, that, after going over to the material picture in the sense of
using the full pull-back of $h_{ij}$, the elastic equation gives us terms
which break diffeomorphism invariance.

\section{}\label{sec:newtonian}
The following Newtonian version of the argument in this paper
was contributed by the anonymous referee.
In the Newtonian theory the equations, with the standard definition of the
stress, taking the specific mass to be 1, are
\begin{subequations} \label{eq:newt-elast}
\begin{align}
\nabla_j \sigma_i{}^j + n \partial_i U &= 0 \quad \text{ in $f^{-1}(\Bo)$},
\label{eq:newt1}
\\
\sigma_i{}^j n_j \big{|}_{\partial f^{-1} (\Bo)} &= 0 , \label{eq:newt2} \\
\Delta U&= 4\pi G n \chi_{f^{-1} (\Bo)} \quad \text{ in $\Re^3_{\Sp}$}
\end{align}
\end{subequations}
The argument to show that the projected version of (\ref{eq:newt1}) implies
(\ref{eq:newt1}) would go as follows. We define $X_i$ to be the solution of
\begin{equation}\label{eq:newt4}
\Delta X_i = \nabla_j \sigma_i{}^j + n \partial_i U
\end{equation}
tending to zero at infinity. Since
$$
\rho_i = \nabla_j \sigma_i{}^j + n \partial_i U
$$
the right hand side of (\ref{eq:newt4}), has compact support, $X_i$ has a
multipole expansion in a neighborhood of infinity,
$$
- 4\pi X_i = \frac{M_i}{r} + \frac{x^j D_{ij}}{r^3} + O(r^{-3})
$$
Here
\begin{align*}
M_i &= \int_{R^3} \rho_i(x) d^3 x \\
D_{ij} &= \int_{R^3} x^j \rho_i(x) d^3 x .
\end{align*}
Writing
$$
n\partial_i U = \nabla_j \Theta_i{}^j
$$
where $\Theta_i{}^j = \Theta_{ij}$ is given by (\ref{theta}) with $h_{ij}$
replaced by $\delta_{ij}$, using the fact that $\Theta_{ij} = O(r^{-4})$ at
infinity, and taking into account (\ref{eq:newt2}), we obtain
\begin{align*}
M &= 0 \\
D_{ij} &= - \int_{\Re^3} (\sigma_i{}^j + \Theta_i{}^j ) d^3 x \\
\intertext{Thus,}
X_i &= - \frac{D_{ij} x^j}{4\pi r^3} + O(r^{-3}),
\end{align*}
where $D_{ij} = D_{ji}$, in agreement with (\ref{eq:Vform}). Lemma
\ref{lem:part-int} then applies with $\Delta$ in the role of $L$ and Lemma
\ref{lem:xiLV-est} applies with $\delta_{ij}$ in the role of
$h_{ij}$. Proposition \ref{prop:VQbound} also applies. The argument of the
proof of Theorem \ref{thm:main} then applies to show that $X_i = 0$,
therefore equation (\ref{eq:newt1}) holds.





\subsection*{Acknowledgements} 
We are grateful to the anonymous referee for reading the paper carefully and
suggesting important improvements. We also thank him 
for providing the material presented
in appendix \ref{sec:bobbyapp}. 

The Newton Institute, where part of the work was done, provided support and
hospitality for which we are grateful. RB thanks the Albert Einstein
Institute for its hospitality. BS would like to thank the members of the geometric
analysis group at the AEI for helpful discussions in the initial stages of
this work. 
LA was supported in part by the National Science Foundation Grant 
DMS-0407732.
RB was supported by Fonds zur F\"orderung der Wissenschaftlichen
Forschung, Project Nr. P16745-N02


\frenchspacing
\bibliographystyle{plain}



%
%
%
%

\end{document}